\documentclass[fleqn,usenatbib]{mnras}

\usepackage{newtxtext,newtxmath}

\usepackage[T1]{fontenc}

\DeclareRobustCommand{\VAN}[3]{#2}
\let\VANthebibliography\thebibliography
\def\thebibliography{\DeclareRobustCommand{\VAN}[3]{##3}\VANthebibliography}

\usepackage{scalerel}
\usepackage{tikz}
\usetikzlibrary{svg.path}

\definecolor{orcidlogocol}{HTML}{A6CE39}
\tikzset{
  orcidlogo/.pic={
    \fill[orcidlogocol] svg{M256,128c0,70.7-57.3,128-128,128C57.3,256,0,198.7,0,128C0,57.3,57.3,0,128,0C198.7,0,256,57.3,256,128z};
    \fill[white] svg{M86.3,186.2H70.9V79.1h15.4v48.4V186.2z}
                 svg{M108.9,79.1h41.6c39.6,0,57,28.3,57,53.6c0,27.5-21.5,53.6-56.8,53.6h-41.8V79.1z M124.3,172.4h24.5c34.9,0,42.9-26.5,42.9-39.7c0-21.5-13.7-39.7-43.7-39.7h-23.7V172.4z}
                 svg{M88.7,56.8c0,5.5-4.5,10.1-10.1,10.1c-5.6,0-10.1-4.6-10.1-10.1c0-5.6,4.5-10.1,10.1-10.1C84.2,46.7,88.7,51.3,88.7,56.8z};
  }
}

\newcommand\orcidicon[1]{\href{https://orcid.org/#1}{\mbox{\scalerel*{
\begin{tikzpicture}[yscale=-1,transform shape]
\pic{orcidlogo};
\end{tikzpicture}
}{|}}}}

\usepackage{hyperref} 

\usepackage{graphicx}	
\usepackage{amsmath}	
\usepackage[export]{adjustbox}
\usepackage{subfig}
\usepackage{fancyhdr}
\usepackage{url}
\usepackage{float}
\usepackage{multirow}
\usepackage{appendix}
\usepackage[portuges]{babel}
\usepackage[utf8]{inputenc}
\usepackage{framed}
\usepackage{tcolorbox}
\usepackage{adjustbox}
\usepackage{stackengine}
\usepackage{framed}
\usepackage{multicol}
\usepackage{setspace}

\title[2001 SN$_{263}$ - the contribution of their irregular shapes on the neighborhood dynamics]
{2001 SN$_{263}$ - the contribution of their irregular shapes on the neighborhood dynamics}

\author[G. Valvano; O. C. Winter; R. Sfair; R. Machado Oliveira; G. Borderes-Motta;]
        {G. Valvano$^{1}$\thanks{E-mail:  giulia.valvano@unesp.br}\orcidicon{0000-0002-7905-1788}\,
        O. C. Winter$^{1}$\thanks{E-mail:  othon.winter@unesp.br}\orcidicon{0000-0002-4901-3289}\,
        R. Sfair$^{1,}$ $^{2}$\thanks{E-mail: rafael.sfair@unesp.br}\orcidicon{0000-0002-4939-013X}
        R. Machado Oliveira$^{1}$\thanks{E-mail: rai.machado@unesp.br }\orcidicon{0000-0002-6875-0508}\,
    \newauthor
        G. Borderes-Motta$^{3}$\thanks{E-mail: gabriel.borderes@uc3m.es}\orcidicon{0000-0002-4680-8414}\
\\
$^{1}$ Grupo de Din\^amica Orbital e Planetologia, S\~ao Paulo State University, UNESP, Guaratinguet\'{a}, CEP 12516-410, 
  S\~{a}o Paulo, Brazil\\
$^{2}$ Institut für Astronomie und Astrophysik, Eberhard Karls Universität Tübingen, Germany\\
  $^{3}$ Bioengineering and Aerospace Engineering Department, Universidad Carlos III de Madrid, Leganés, 28911, Madrid, Spain} 
\date{Accepted XXX. Received YYY; in original form ZZZ}

\pubyear{2022}

\begin{document}
\label{firstpage}
\pagerange{\pageref{firstpage}--\pageref{lastpage}}
\maketitle

\begin{abstract}
The first proposed Brazilian mission to deep space, the ASTER mission, has the triple asteroid system (153591) 2001 SN$_{263}$ as a target. One of the mission's main goals is to analyze the physical and dynamical structures of the system to understand its origin and evolution. The present work aims to analyze how the asteroid's irregular shape interferes with the stability around the system. The results show that the irregular shape of the bodies plays an important role in the dynamics nearby the system. For instance, the perturbation due to the (153591) 2001 SN$_{263}$ Alpha's shape affects the stability in the (153591) 2001 SN$_{263}$ Gamma's vicinity. Similarly, the (153591) 2001 SN$_{263}$ Beta's irregularity causes a significant instability in its nearby environment. As expected, the prograde case is the most unstable, while the retrograde scenario presents more stability. Additionally, we investigate how the solar radiation pressure perturbs particles of different sizes orbiting the triple system. We found that particles with a 10-50 cm radius could survive the radiation pressure for the retrograde case. Meanwhile, to resist solar radiation, the particles in prograde orbit must be larger than the particles in retrograde orbits, at least one order of magnitude.
\end{abstract}


\section{Introduction}
Over the past years, several space missions have targeted asteroids to comprehend their
formation, evolution, and dynamics since they are the remaining bodies of the inner Solar System. The
NEAR-Shoemaker mission was the first spacecraft to orbit the asteroid (433) Eros and ended up touching its
surface, being also the first one to land on an asteroid \citep{veverka2000near, prockter2002near}. The
asteroid (25143) Itokawa was imaged by the Hayabusa spacecraft and was the first mission to collect samples 
from the surface of an asteroid \citep{fujiwara2006rubble, yoshikawa2015hayabusa}. Similarly, the Hayabusa 2 spacecraft
performed a touchdown in 2019 on the asteroid (162173) Ryugu, and the  collected samples 
arrived on Earth in December 2020 \citep{kawaguchi2008hayabusa, muller2017hayabusa}. The OSIRIS-REx mission
also plans to return in 2023 a sample collected from 
the asteroid (101955) Bennu \citep{lauretta2015osiris, lauretta2017osiris}, another spinning-top shape
asteroid as Ryugu. The first space mission intended  to demonstrate an asteroid deflection by kinetic
impactor is DART (Double Asteroid Redirection Test), and the mission target is the binary near-Earth 
system Didymos-Dimorphos \citep{cheng2016asteroid}. The multiple system of asteroids is notably 
attractive for space missions as they can provide a manifold of observational possibilities, investigation
of the system origin, and the dynamics involved. Based on these same concepts, the ASTER mission team
chose as its target a
triple system to be the first Brazilian deep space mission \citep{su2010}. 

The target of the ASTER mission, the asteroid (153591) 2001 SN$_{263}$, 
was discovered in 2001 by LINEAR (Lincoln Near-Earth Asteroid Research). 
However, radar observations from Arecibo in 2008 revealed 
the asteroid is, in fact, a triple system. 
The system is composed of the central body called Alpha and two 
secondary bodies, Beta and Gamma, being Gamma the smallest. Using the same data, \citet{Fang2011} 
made the best fit for their physical and orbital parameters. 
\citet{Araujo2012, Araujo2015}, motivated by the ASTER mission announcement, 
investigated the stability of internal and external regions to the triple system 
for prograde and retrograde orbits. They made a set of numerical simulations 
considering the asteroids as spherical bodies and reported regions of stability 
between them. \citet{prado2014mapping} plotted the perturbation map for a spacecraft orbiting the triple system.
Their results 
showed that Gamma causes a more significant perturbation in the system, 
one order of magnitude larger than 
the perturbation caused by Beta and two orders larger than Alpha's oblateness for particles with a semi-major axis around 2-40 km.
\citet{sanchez2019searching} also searched for orbits around the system but 
considering the solar radiation pressure. They mapped the duration of the orbits 
considering values of semi-major axis, eccentricity, and 
inclination. They identified some possibly stable areas and a limit on the size of the spacecraft
solar panels for their stable results. 
In none of these previous works the actual irregular shape of the bodies was taken into account.
Adopting  the polyhedral shape model for each asteroid, derived by \citet{Becker2015}, \citet{winter2020asteroid} 
explored the geometric and geopotential 
topographic of the bodies and the dynamical environment around each 
system component. They determined 12 equilibrium points for the central body. Also, they found that the particles initially in a cloud around Alpha would fall preferentially on the equatorial bulge and the polar regions.

Considering the singular dynamics and the previous analyses, 
we aim to investigate the stability of different regions of the system, taking into 
account the gravitational effects due to irregular shape models 
of the bodies and  the effects the solar radiation 
pressure as an extension of the results of \citet{Araujo2012, Araujo2015}. First, in section \ref{triple}, we 
present the physical and orbital properties of the system. Then, a set of numerical simulations of discs of test particles in the triple system  is presented and discussed in section 
\ref{nearby}. In the next section, we add the perturbation of the solar radiation pressure and 
investigate how this perturbation changes the stability of particles in the system. Next, long-period simulations are 
presented in the section \ref{longo}. Then, we provide the final comments for our results (section \ref{final}).


\section{The triple asteroid system (153591) 2001 SN$_{263}$}
\label{triple}

In this section, we discuss about the orbital and physical properties, and the irregular shape models of the (153591) 2001 SN$_{263}$ bodies. 

\begin{table*}
\centering
\caption{Physical and orbital data of each component of the triple asteroid system (153591) 2001 SN$_{263}$}
\label{table: infos} 
\begin{tabular}{c|c|c|c|c|c|c|c|c|c}
\hline\hline
Body & Volumetric & Density$^{b}$ & Spin$^{b}$ & Mass$^{a}$ & Orbit & Semi-major & Eccentricity$^{a}$ & Inclination$^{a}$ & Orbital\\
  & radius (km) & (g$\cdot$cm$^{-3}$) & (hours) & (10$^{10}$ kg) & & axis $^{a}$ &  & (deg) & period\\
\hline
\hline
Alpha & 1.25 & 1.10 & 3.4256 & 917.466 $\pm$ 2.235 & Sun & 1.99 au & 0.48  & 6.7  & 2.8 years$^{c}$\\
Beta & 0.39 & 1.00 & 13.43 & 24.039 $\pm$ 7.531 & Alpha & 16.63 km & 0.015 & 0.0 & 6.23 days$^{a}$\\
Gamma & 0.22 & 2.30 & 16.40 & 9.773 $\pm$ 3.273 & Alpha &3.80 km & 0.016 & 14 & 16.46 hours$^{a}$\\ 
\hline\hline
\end{tabular}
  \begin{flushleft}
  	\quad  {\footnotesize $^{a}$ \citet{Fang2011}}
  				
  	\quad  {\footnotesize $^{b}$ \citet{Becker2015}}
  				
  	\quad  {\footnotesize $^{c}$ JPL. Website: https://ssd.jpl.nasa.gov/}
\end{flushleft}
\end{table*}


\subsection{Orbital properties}
\label{orbital}

\begin{figure*}
\begin{center}
\subfloat{\includegraphics*[trim = 2cm 5cm 0mm 0mm, width=2\columnwidth]{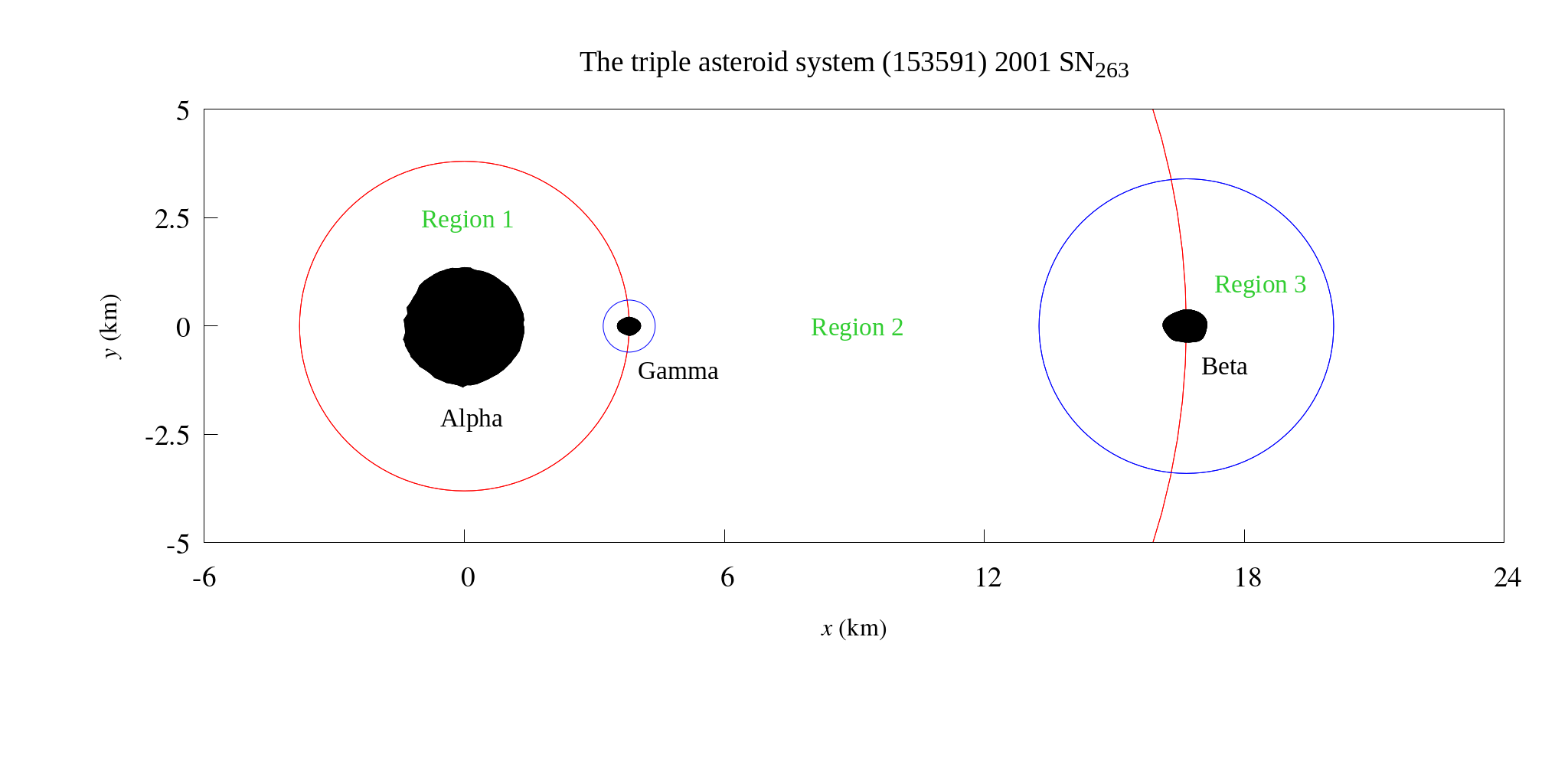}}
\end{center}
\caption{Representation of the triple system (153591) 2001 SN$_{263}$ in the Alpha centered equatorial plane, $xoy$. The red circles represent the orbits of Gamma and Beta and, also, the collision lines for regions 1 and 2. The blue circles represent the Hill's radii of Beta-Alpha (ejection distance for region 3) and Gamma-Alpha.}
\label{fig:orbit}
\end{figure*}

On February 20$^{\textrm th}$ of 2008, the triple system 2001 SN$_{263}$ reached a distance of 0.066 au 
from the Earth and was observed by Arecibo and Goldstone radio telescopes. 
Based on radar data, \citet {Fang2011} provided orbital fits for the system 
and the orbital evolution considering the mutual gravitational perturbation 
among the asteroids \citep{Fang2011}. The semi-major axes, eccentricities, and inclinations found are given in Table 
\ref{table: infos}, and a representation of the system in the Alpha centered equatorial 
plane is shown in Fig. \ref{fig:orbit}. The red circles in Fig. \ref{fig:orbit} 
represent the orbits of the secondary bodies and the blue circles their respective Hill's radii. 
The Hill's radii can be used as a parameter to bound the regions where a body is gravitationally dominant. 
We computed the Hill's radii of Beta and Gamma with respect to Alpha, 
which are $\sim$3.4 km and $\sim$ 600 m, respectively. 
Then, we delimit the regions that will be used in our analysis in the same way as was done by \citet{Araujo2012, Araujo2015}, since we will compare the stability by regions. 
Region 1 is delimited as the region interior to  Gamma's orbit; region 2 is between Gamma's 
and Beta's orbits and region 3 is just delimited as the region interior to  Beta's Hill's radius.

Besides the orbital elements, \citet{Fang2011} present fits for the 
pole direction of Beta and Gamma. 
They stated that about 25\% of their adjusts resulted in retrograde orbits 
(contrary to Alpha's spin). However, they reported that Beta and Gamma are more likely to be prograde. 
The retrograde result was possible due to a wide range of 
spin axis orientations for Alpha, 
though the orbital orientations of the secondary bodies are well defined. 
\citet{Becker2015} also investigated the 2001 SN$_{263}$ orientation and searched 
for the best fit for the spin axis of Alpha. 
They reported an obliquity of about 166$^{\circ}$ and the spin-axis pole of 
($\lambda, \beta$) = (309, $-$80)$\pm$15$^{\circ}$, which is similar to 
the pole orientation reported by \citet{Fang2011}. 
To reproduce a similar Alpha's pole orientation, we considered, for simplicity, 
the obliquity of Alpha and 270$^{\circ}$ for the pole's longitude.


\subsection{Shape model and physical properties}
\label{shape}

The 2001 SN$_{263}$ shape model was derived by \citet{Becker2015} from the 
combination of radar and light curves data taken in 2008 from Arecibo and 
eight other observatories. The three shape models have equivalent diameters of 
about 2.5 km, 0.77 km, and 0.43 km for Alpha, Beta, and Gamma, respectively. 
A polyhedron modeled each body with 1148 vertices and 2292 triangular 
faces. In the same study, \citet{Becker2015} derived a rotational period 
for each body (Table \ref{table: infos}). 

\citet{Fang2011} also reported the mass for each asteroid (Table \ref{table: infos}). 
Considering the mass and the volume of the shape models, \citet{Becker2015} 
estimated a bulk density for the bodies and concluded that the bodies are 
carbonaceous asteroids of type B. 
\citet{perna2014triple} came to the same conclusion for the B-type classification by comparing the reflectance spectra with laboratory spectra data. They also concluded that the bodies have an organic- and magnetite-rich surface composition, similar to the CI carbonaceous chondrites. 
The bulk density of Alpha and Beta are similar and compatible with the BCG-types 
[B-, C-, Cb- and Cg-type] range of about 0.7-1.7 g$\cdot$cm$^{-3}$ 
\citep{marchis2012multiple, vernazza2015interplanetary}. 
However, the Gamma's bulk density is much higher. 
The grain density of carbonaceous chondrites is approximately 2.42-5.66 g$\cdot$cm$^{-3}$, 
with an average of 3.44 g$\cdot$cm$^{-3}$ 
\citep{macke2011density, carry2012density, flynn2018physical, ostrowski2019physical}. 
Thus, considering the mean value, Gamma must have a porosity smaller than Alpha and 
Beta if its bulk density is 2.3 g$\cdot$cm$^{-3}$. 
This difference may indicate that Gamma could have been originated from a different parental body. 
If we consider a scenario with a collision between two different parent bodies with different 
bulk densities \citep{michel2020collisional}. 
We also have to consider the spinning-top shape of the central body. 
After the disruption collision, the aggregate parental body of Alpha may have started to YORP spin-up, 
resulting in the current diamond shape  
\citep{rubincam2000radiative, walsh2008rotational, sanchez2016disruption, walsh2018rubble}.

\begin{figure}
\begin{center}
\subfloat{\includegraphics*[trim = 0mm -0.3cm 0mm 0mm, width=1\columnwidth, frame]{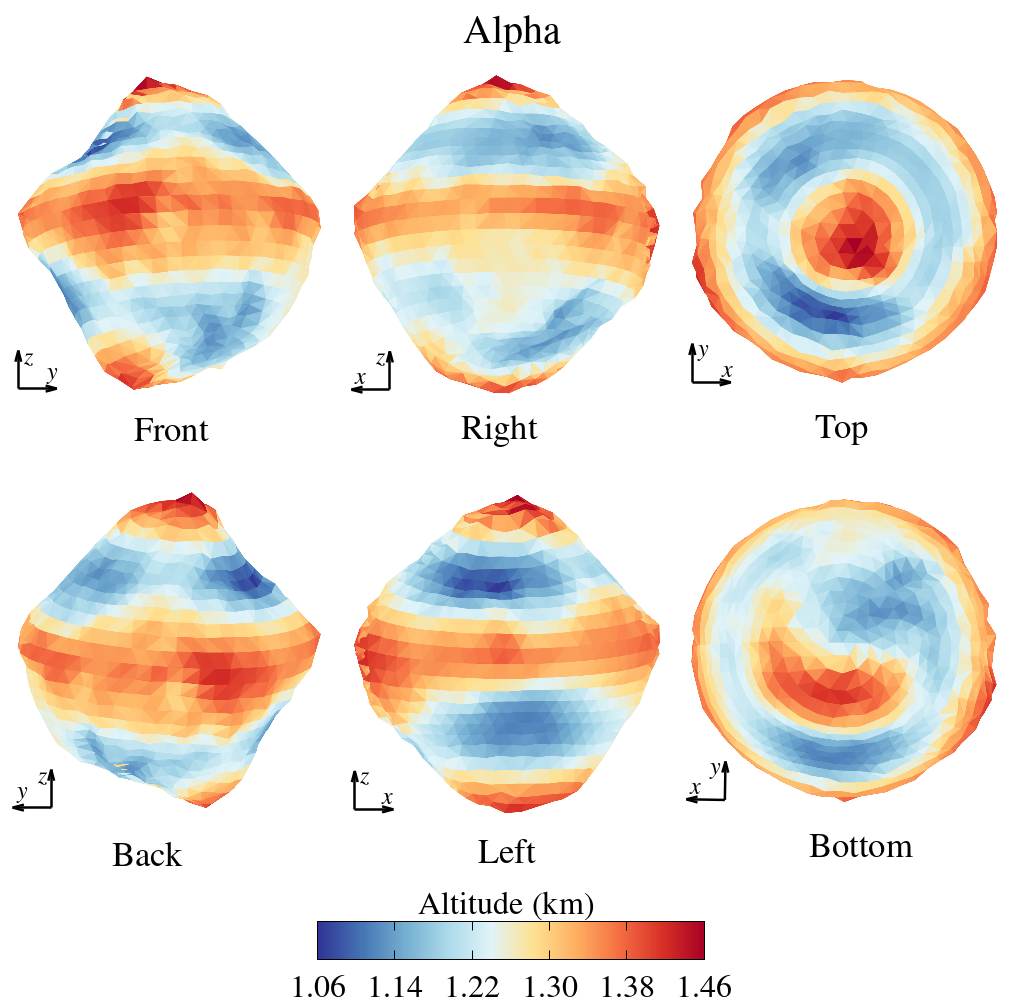}}\\
\subfloat{\includegraphics*[trim = 0mm 7cm 0mm 0mm, width=1\columnwidth, frame]{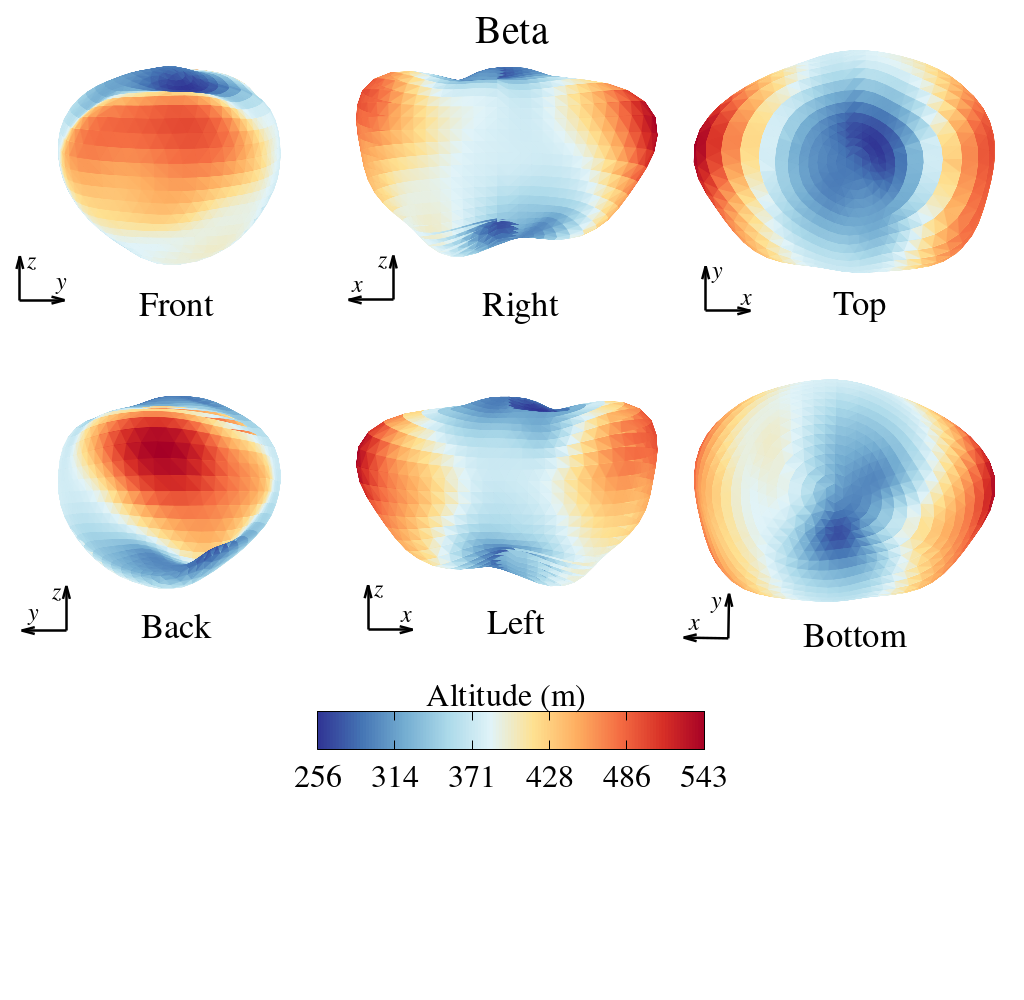}} \\
\subfloat{\includegraphics*[trim = 0mm 5.3cm 0mm 0mm, width=1\columnwidth, frame]{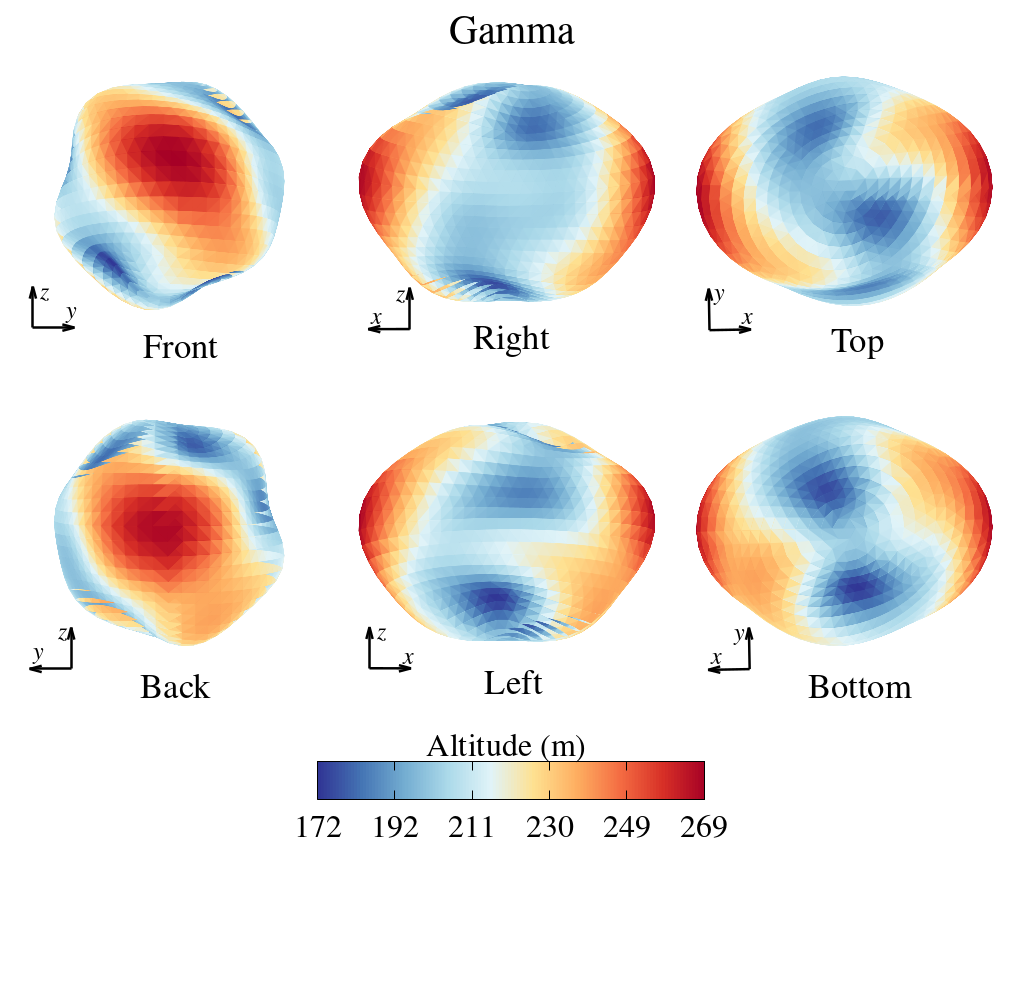}}
\end{center}
\caption{Map of the geometric altitude across the surfaces of Alpha, Beta, and Gamma under different views.}
\label{fig:geom}
\end{figure}

This diamond shape has larger values of altitude on its equatorial bulge and the polar regions (Fig. \ref{fig:geom}).
We also can see that the mid-latitude regions have lower altitudes. 
Alpha's shape is characteristic of spinning-top asteroids that are a result of their fast spin \citep{harris2009shapes, hirabayashi2020spin}. 
Differently from Alpha, Beta has flattened poles with the lower altitudes 
located in these regions (see Top and Bottom views in Fig. \ref{fig:geom} – Beta). 
The maximum altitudes for this body are in the equatorial extremity. Gamma also has the upper altitudes  
in its equatorial extremities. However, its poles are not so flat as Beta's ones.
We can see in both poles the existence of moderate altitude followed by valleys 
with lower values (Top and Bottom views of Fig. \ref{fig:geom} – Gamma). 
Note that Beta has the more considerable relative variation of altitudes, with the maximum reaching about twice the
minimum altitude. On the other hand, Alpha and Gamma have a similar relative variation, the ratio between the maximum and
minimum altitudes of Alpha is about 1.4, and Gamma is 1.6.  

The irregularities in the shapes of the bodies  
influence their gravitational field. Thus, if we treat the system as a point of mass (or spherical bodies), 
as done by \citet{Araujo2012, Araujo2015}, it may not be a satisfactory approximation.
Using the volumetric radius of Alpha (Table \ref{table: infos}), 
we will lose peaks of at least 210 m of altitude. For Beta, we will miss about 150 m and Gamma 50 m and, 
consequently, their irregularities (Fig. \ref{fig:geom}). 
So, to identify the contribution due to the irregular format of the system's bodies, 
in the following sessions, we will present  a set of numerical simulations with test particles  
in the triple system gravitational field to analyze their stability.


\section{Effects on the nearby environment}
\label{nearby}

The dynamics nearby the system were studied by \citet{Araujo2012, Araujo2015} to identify possible
stable regions for a spacecraft or locations for natural objects, such as fragments or 
dust.
Then, they explored the stability of massless particles discs in
prograde and retrograde trajectories in the system 
through a set of numerical simulations considering the triple system, the Sun, Earth,  Mars, and
Jupiter as a point of mass (or spherical objects) and included the oblateness of Alpha ($J_2$).
The particles were distributed in different regions: between Alpha and Gamma, Gamma and
Beta, around Beta, and an outer region with the particles orbiting the triple system. 

In the current work, our goal is to advance the results of \citet{Araujo2012, Araujo2015} considering
a more realistic scenario. Our main concerns are the effects caused by the irregular shapes 
of the triple system components on the stability of the particles. 
Another  perturbation that we will consider is the solar radiation pressure, which  might be relevant
depending on the particle size.

Thus, we performed a set of numerical simulations 
of test particles  using the 
\textsc{N-BoM} package \citep{winter2020asteroid}. 
Mascon models computed the gravitational potential of the irregular bodies, the mass of the body is uniformly divided into a tridimensional equally spaced grid \citep{Geissler1996,borderes-motta18}. We modeled Alpha using a with 24328 mass points and Beta with 24065-grid-point.
Since we want to verify how the irregular shape of the bodies interferes with the particle 
stability, we first reproduced the figures of \citet{Araujo2012, Araujo2015} to make the comparisons. 
For simplicity, we did not consider the perturbations of the planets 
since they do not contribute to the stabilization of the particles. 
Hereinafter, ``standard case" refers to the simulations based on \citet{Araujo2012, Araujo2015}, while the ``mascons case" is our 
solution that takes into account the shape models.

The distribution of the initial conditions of particles was made in pairs of
semi-major axis and eccentricity.
For regions 1 and 2, the particles were orbiting Alpha with 
semi-major axes in the intervals [1.4, 3.2] km and [4.5, 13.5] km, respectively, while 
the particles in region 3 were orbiting Beta with semi-major axes between 0.8 km and 3.4 km. 
Each range of the semi-major axis 
was evenly divided into steps of 200 m, and the eccentricity ranged from
0.0 to 0.5 in steps of 0.05. The inclination for the particles was initially 0$^{\circ}$ for prograde trajectories and 180$^{\circ}$ for retrograde trajectories.
We consider prograde trajectories the orbits that have the same direction of the angular momentum of the orbited body.
In each case, the initial conditions were randomly chosen for the other angular orbital elements.

For each region, we considered a different ejection
distance. The ejection distance means that the particle has left its initial region and does not
necessarily leave the system. The ejection distance for region 1 was 3.80 km, which corresponds to
the semi-major axis of Gamma. Region 2 has an ejection distance of 16.63 km, corresponding to the
semi-major axis Beta (Table \ref{table: infos}); the Hill's radii of Beta, 3.4 km, delimits
the ejection distance for region 3. The system was integrated for 2 years, and we classified the stability according to the percentage of particles for a given pair $(a,e)$ 
that survived (i.e., were not ejected nor collided with a body) 
the entire time.

\citet{Araujo2012, Araujo2015} also considered a fourth region, external to the triple
system, starting from 20 km. However, we did not simulate this region since it is far distant
from bodies Alpha and Beta. Therefore, their irregular shape will not interfere significantly with the dynamics of the particles' stability.

\subsection{Orbits of the triple system components}
 To study the dynamics of the triple system, we considered Alpha at the system origin and the
orbital elements of Beta and Gamma provided by \citet{Fang2011} (Table \ref{table: infos}), 
with the other angular elements set as zero. 
\citet{Fang2011} derived the orbital periods of Beta and Gamma by performing simulations 
considering the triple system and Alpha’s 
$J_{2}$ gravitational harmonic, and reported orbital periods 
of about 149.4 hours for Beta and 16.46 hours for Gamma.
However, using the mentioned initial conditions for our simulations, 
the orbital
period of Beta diverged from their reported orbital period. 
We used the Lomb-Scargle periodogram \citep{1976Ap&SS..39..447L, 1982ApJ...263..835S} and the searching
ranges of [97, 200] and [10, 20] hours for Beta and Gamma orbital periods, respectively. 
The output distance from Alpha and time were used to fit the periodogram model. 
The period considered for the periodogram calculations was the 2-year integration
and we obtained an orbital period of about 136.26 hours for Beta when the three bodies were considered.
However, when the orbital period is simply calculated using the third Kepler law,
it matches the reported value \citep{Fang2011}. 
On the other hand, despite of the divergence of Beta's orbital period, the orbital
period of Gamma did not diverge when we considered a three-body problem. 
This may be related to the fact that Gamma is closer to Alpha than Beta,
thus Beta does not directly affect Gamma's orbit significantly. 
Nevertheless, according to \citet{prado2014mapping} perturbation's map,
the perturbation of Gamma is one order of magnitude larger than
Beta's, and Gamma's perturbation is about ten times stronger for
test particles orbiting Alpha, but close to Beta, thus Beta  experiences some perturbation due to Gamma.

\begin{figure}
\begin{center}
\includegraphics*[trim = 0mm -0.3cm 0mm 0mm, width=1\columnwidth]{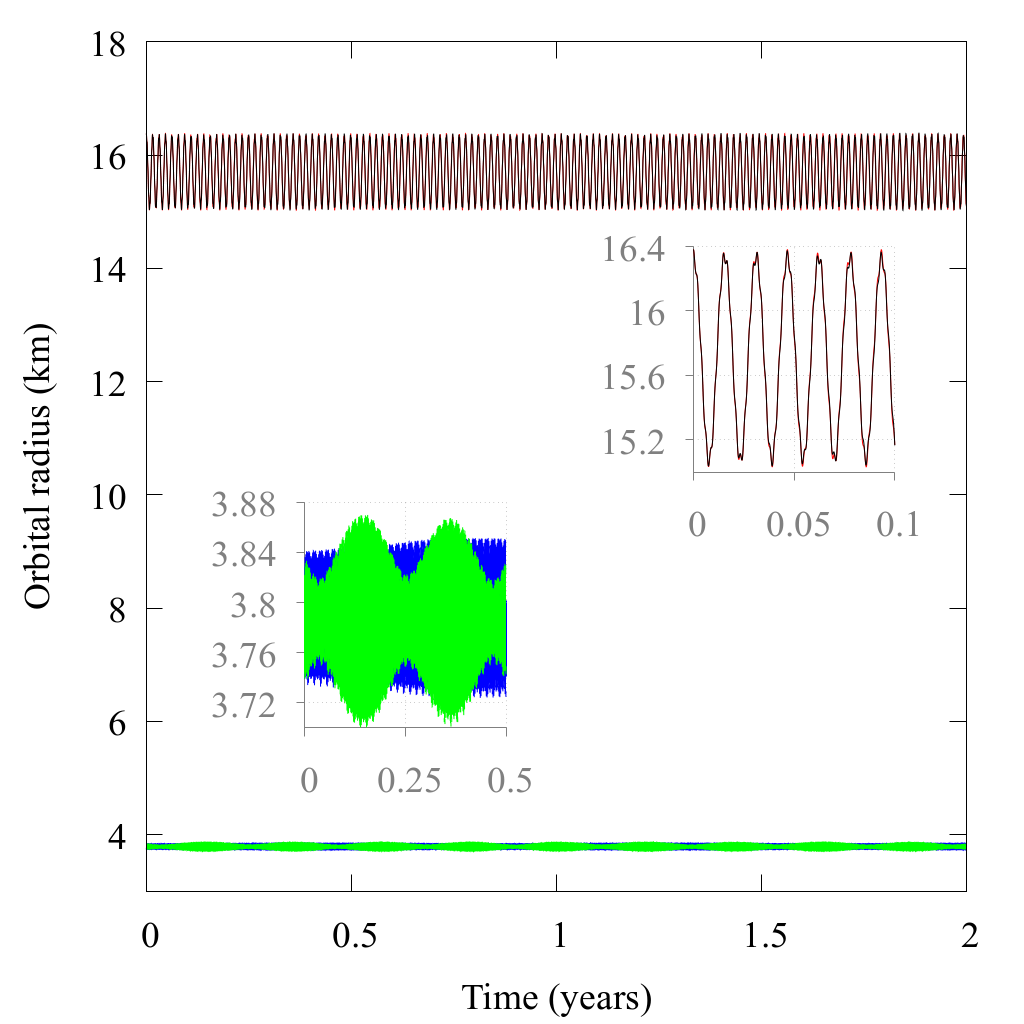}
\end{center}
\caption{Orbital radius variation across time of Beta and Gamma for two years. Considering the Alpha as a spherical body with the perturbation of its $J_{2}$ gravitational harmonic, the red and blue lines represent the orbital variation of Beta and Gamma, respectively. The black and green lines showed the orbital variation of Beta and Gamma, respectively, for the Alpha's irregular case scenario. The zoom in the figure shows the orbital radius of Gamma (inferior) and Beta (superior) for a period of six months.}
\label{fig:zoom}
\end{figure}

In order to measure how much the irregular shape of Alpha affects the orbits of Beta and Gamma, 
we simulated two scenarios: the standard case with Alpha as an oblate object (i.e. with the $J_{2}$ coefficient)
and other with the irregular shape of Alpha (mascons case). 
Due to the complexity of modeling the mutual interaction between two irregularly shaped bodies, in our model we considered that the mascons of Alpha affect Beta, while the disturbance caused by Beta on Alpha corresponds to a point of mass interaction
From Fig. \ref{fig:zoom} we see that the orbital radius 
of Beta does not suffer a change in its amplitude of oscillation due to  Alpha's irregular shape. On the other hand, a different behavior of the orbital radius of Gamma is noticed 
when the irregular shape  is considered. The peak to peak amplitude of the orbital
radius of Gamma in the standard case is about 40 m smaller than in the
mascons simulation. The oscillation frequency of the orbital radius is also different,
being 3.5 times larger for the mascons case when compared with the oblate model. 
Thus, the Alpha irregular shape  does play an
important role in the dynamics of Gamma, but does not  significantly change the
orbit of Beta.

We also ran a simulation to verify the perturbation of the irregular shape model of Beta, and 
no significant perturbation is noted in the orbit of Gamma.
In summary, the main perturbation is caused by the irregular shape model of Alpha on the trajectory of Gamma.

\subsection{Region 1}
\label{region1}

\begin{figure}
\begin{center}
\subfloat[]{\includegraphics*[trim = 0mm 0cm 0cm 0mm,
width=0.9\columnwidth]{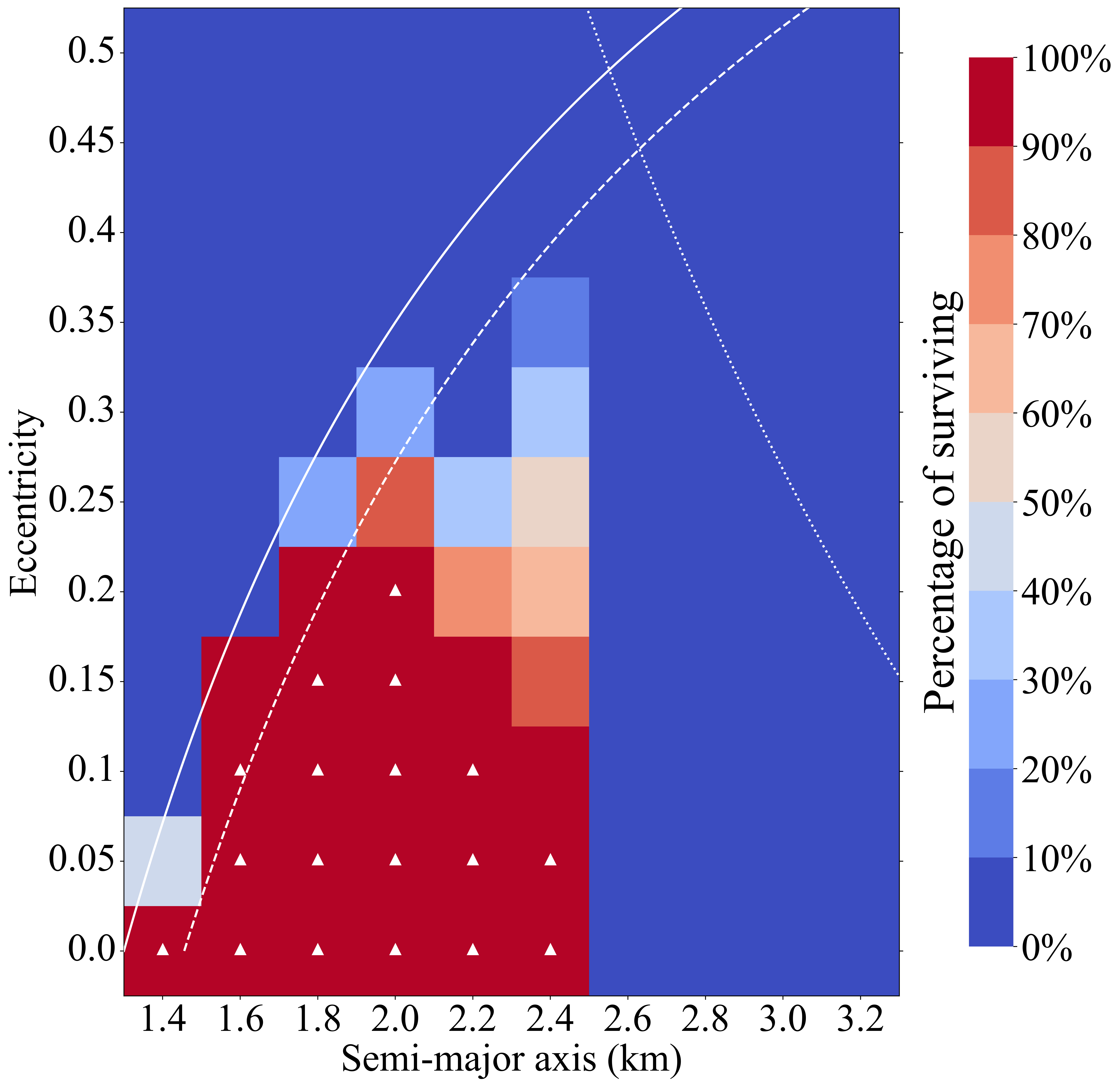}\label{afig:Reg1_SemSrp}}
\end{center}
\caption{\label{fig:Reg1_SemSrpP} Diagram of stability for region 1 for two years considering prograde trajectories. The diagram is a reproduction of figure 2a from the standard case \citep{Araujo2012} with the triple system as points of mass and the $J_2$ of Alpha. Each box represents a set of 100 particles and the color bar represents the percentage of surviving particles. The white triangles indicate 100\% of survival. The lines indicate the collision-lines with Alpha (full line for a radius of 1.3 km, dashed line for a radius of 1.46 km), and Gamma (dotted line, corresponding to the ejection distance).}
\end{figure}

Given the initial conditions presented previously for region 1, the combination of semi-major axes
and eccentricities resulted in 11,000 particles since they were considered 100 different particles for each
($a, e$) pair.  Figure \ref{fig:Reg1_SemSrpP} 
is a reproduction  of Figure 2a from 
\citet{Araujo2012}. 
It shows the stability diagram for the prograde
particles and is colored according to the surviving box percentage.  The box with a
white triangle indicates 100\% of survival, and each box represents 100 particles.  The full white
line represents the collision line with Alpha for a radius of 1.3 km (approximately the
Alpha volumetric radius is the same radius adopted by \citet{Araujo2012, Araujo2015}).  The
dashed line is the collision-line for a radius of 1.46 km (corresponding to Alpha upper altitude
calculated on the body's surface, see Fig. \ref{fig:geom}).  The dotted line represents the collision-line
with Gamma.  To compute these limits, we consider 
that the pericentre of the particle's orbit can not be smaller or equal to Alpha's radius, and the apocentre can not be larger
or equal to Gamma's orbit.
Hence, for the collision-line with Alpha we adopted $a \leq R_{Alpha}/(1 - e)$ and for Gamma's, $a \geq a_{Gamma}/(1 + e)$, where $R_{Alpha}$ and $a_{Gamma}$ are, respectively,
the Alpha's radius and Gamma's semi-major axis.

In our simulations, we took into account only the irregular shape of Alpha.  Since Gamma is much smaller (about a hundred times), its irregular shape will not significantly influence particles of region 1.  The irregular shape of Beta was also not considered as its location is too far from
region 1.  For this prograde
simulation, the results revealed that no particle survived for two years. On the other hand, a massive quantity of particles was ejected or
collided within just seven days. 
The initial conditions with a semi-major axis higher than 2.4 km, near Gamma's orbit, were
already unstable in \citet{Araujo2012} study.  However, for values smaller or equal to 2.4 km, near Alpha's surface, there were many stable trajectories with $e\leq 0.25$ (Figure \ref{fig:Reg1_SemSrpP}). As we consider Alpha's shape, the irregularities destabilize all the remaining trajectories of region 1.

\begin{figure}
\begin{center}
\subfloat[]{\includegraphics*[trim = 0mm 0cm 0cm 0cm,
width=0.9\columnwidth]{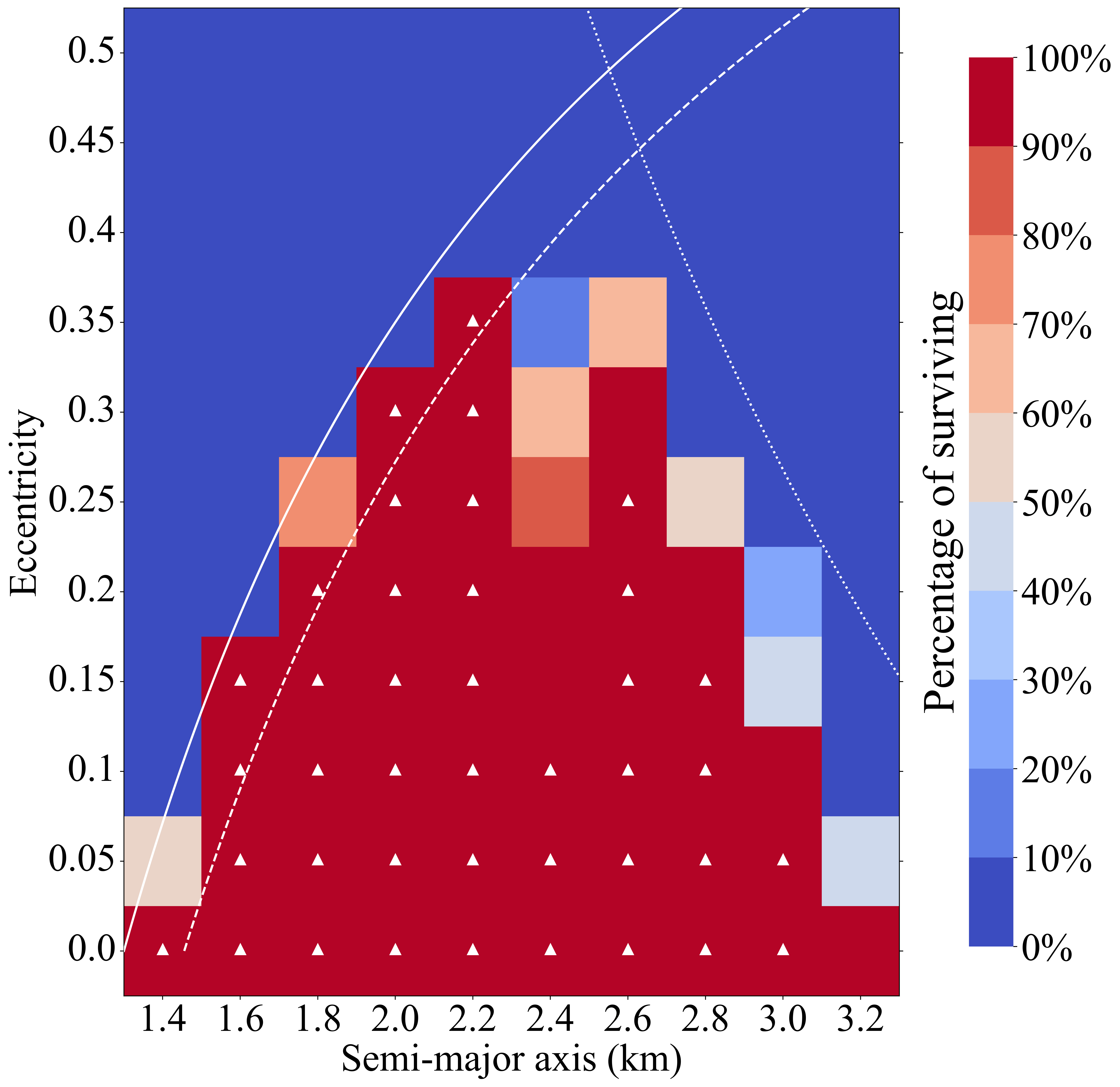}\label{afig:Reg1_SemSrp_Retro}}\\
\subfloat[]{\vspace{3cm}\includegraphics*[trim = 0cm 0cm 0cm 0mm,
width=0.9\columnwidth]{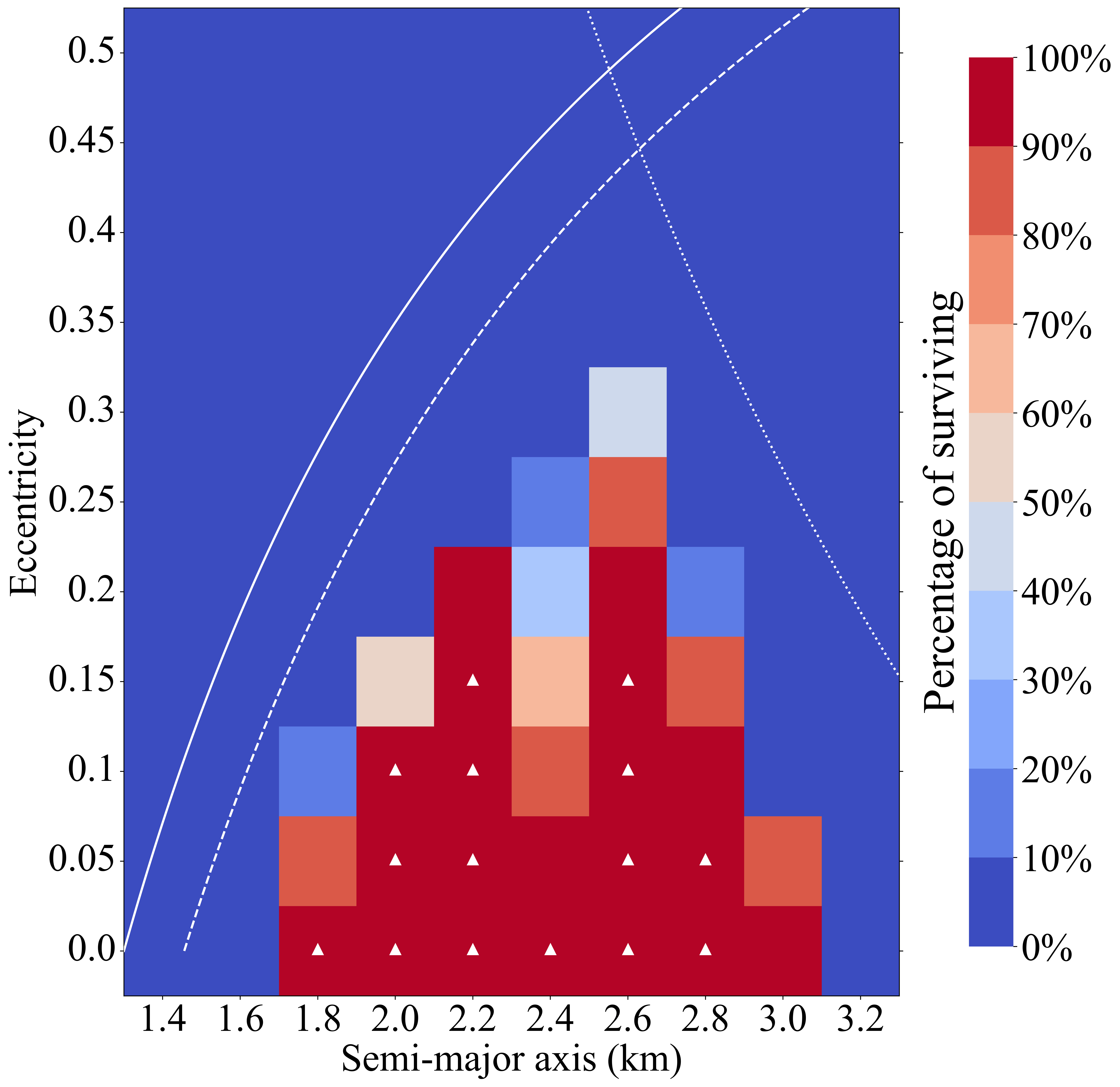}\label{bfig:Reg1_SemSrp_Retro}}
\end{center}
\caption{\label{fig:Reg1_SemSrp_Retro} Region 1 stability diagram for two years considering retrograde trajectories. 
(a) it is a reproduction of figure 2f from standard case \citep{Araujo2015} and (b) shows the system considering the irregular shape of 
Alpha. Each box represents a set of 100 particles, and the color bar represents the percentage of surviving particles. 
The white triangles indicate 100\% of survival. The lines indicate the collision with Alpha 
(full line for a radius of 1.3 km, dashed line for a radius of 1.46 km), and Gamma (dotted line corresponding to the ejection distance).}
\end{figure}

When retrograde particles are examined, an increase of survived particles
occurs (Fig. \ref{fig:Reg1_SemSrp_Retro}). For the standard case,
about 48\% of the particles survived for two years of integration and,  different
from the prograde case (Fig. \ref{fig:Reg1_SemSrpP}), the survived particles can reach a semi-major
axis larger than 2.4 km and also larger eccentricities (Fig. \ref{fig:Reg1_SemSrp_Retro}a). However, the
eccentricity did not exceed values larger than 0.4 for most survived particles. An interesting
feature is a gap in the semi-major axis at 2.4 km. The particles with this
semi-major axis have a 2:1 commensurability of mean motion with Gamma. The particles nearby this
region suffer from a perturbation due to this resonance, and their eccentricity increase in such a way
that they cross the collision line of the region \citep{Araujo2015}. 

A decrease in the survived particles number and
the ranges of semi-major axis and eccentricity was noticed when we consider the mascons case
(Fig. \ref{fig:Reg1_SemSrp_Retro}b). The irregularities significantly affected the particles with pericentre closer to the
central body. 
The gap in the 2.4 km semi-major axis is still present due to the 2:1
commensurability of mean motion with Gamma. This implies a resonance that also causes instability, and due to Alpha's irregular shape, this instability affects particles with initial eccentricities smaller than in the case of Fig. \ref{fig:Reg1_SemSrp_Retro}a. Since their orbits can not reach larger values of
eccentricities without colliding with Alpha or crossing the collision-line of Gamma, the boxes of initially
almost circular orbits have a more significant percentage of survivors (Fig. \ref{fig:Reg1_SemSrp_Retro}b).


\subsection{Region 2}
\label{region2}

\begin{figure*}
\begin{center}
\subfloat[]{\includegraphics*[trim = 0mm 0cm 0cm 0mm,
width=2\columnwidth]{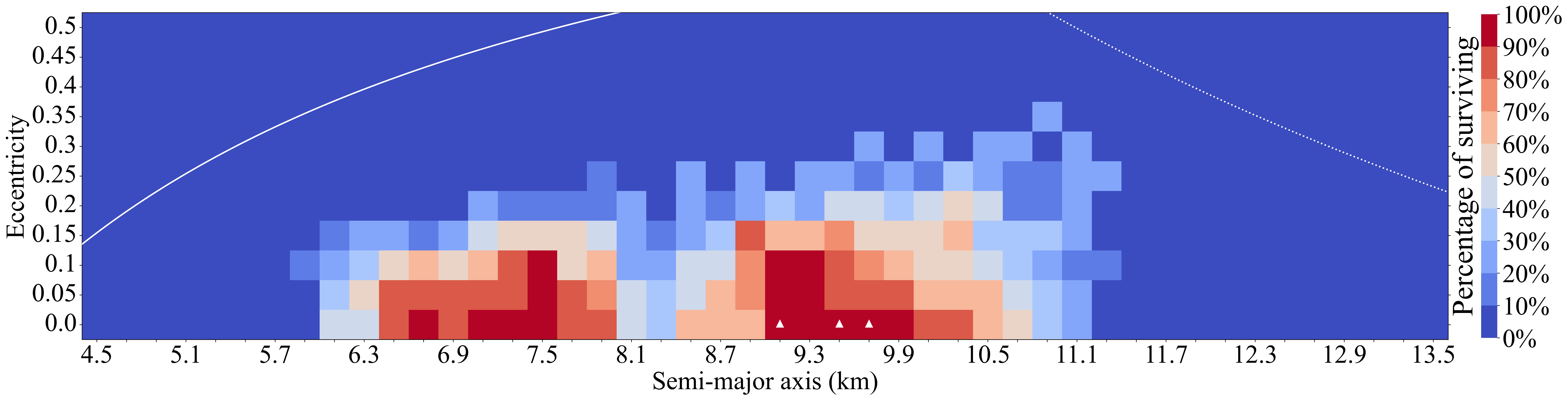}\label{afig:Reg2_SemSrp}}\\
\subfloat[]{\includegraphics*[trim = 0mm 0cm 0cm 0mm,
width=2\columnwidth]{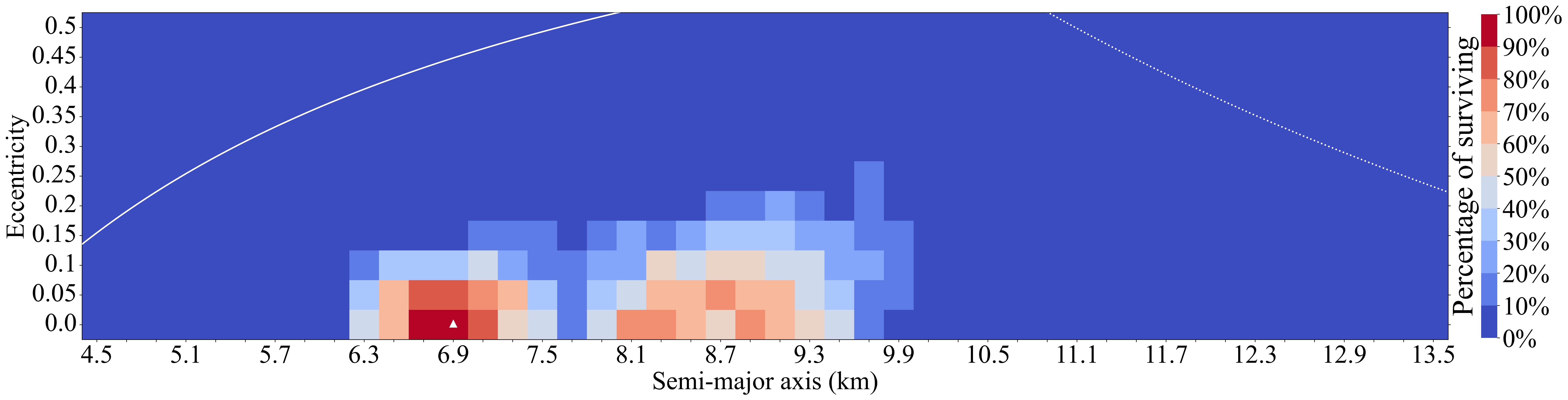}\label{bfig:Reg2_SemSrp}}
\end{center}
\caption{\label{fig:Reg2_SemSrp} Diagram of stability for region 2 for two years considering prograde trajectories. 
(a) it is a reproduction of figure 3a from the standard case \citep{Araujo2012} and (b) shows the system considering the irregular 
shapes of Alpha and Beta. Each box represents a set of 100 particles, and the color bar represents the percentage of surviving particles. 
The white triangles indicate 100\% of survival. The lines indicate the collision lines with Gamma (full line) and Beta 
(dotted line).}
\end{figure*}

The distribution of initial conditions in region 2 resulted in 50,600 particles between Gamma and Beta, where the orbits of
Gamma and Beta correspond to collision lines that delimit the ejection of region 2. To calculate these lines, we assume
that the pericentre of the particle's orbit cannot be smaller or equal to Gamma's semi-major axis
and the apocentre larger or equal to Beta's semi-major axis. Figure \ref{fig:Reg2_SemSrp} shows the diagrams of
stability of  region 2 for the prograde case. 
For our simulations, we consider the irregular shapes of Alpha and Beta. 
A similar structure to the standard case (Fig. \ref{fig:Reg2_SemSrp}) still remains. 
The survived particles are also located in two preferential spots, but now with semi-major axes near 6.9 km and 8.7 km, while for the standard case, they were concentrated around the semi-major axis of 7.3 km and 9.5 km.
The number of survivors decreased by about 67\% compared to the standard case. 
The elongated shape of Beta causes the removal of particles with a semi-major axis of 9.5 km, and a gap was formed at 7.7 km (Fig. \ref{fig:Reg2_SemSrp}b), 
dislocating in about 600 m from the gap formed in the standard case (Fig. \ref{fig:Reg2_SemSrp}a). 
These gaps are associated with a 3:1 mean motion resonance with Beta. Such resonance forces an increase in the eccentricities of the nearby particles 
 destabilizing them and generating the gap.

\begin{figure*}
\begin{center}
\subfloat[]{\includegraphics*[trim = 0mm 0cm 0cm 0mm,
width=2\columnwidth]{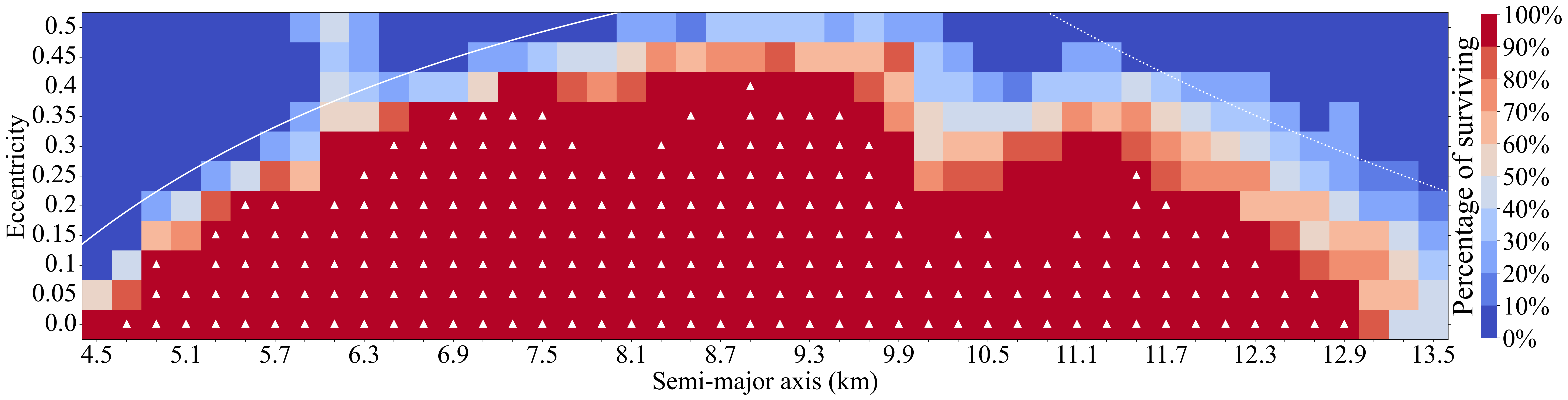}\label{afig:Reg2_SemSrp_Retro}}\\
\subfloat[]{\includegraphics*[trim = 0mm 0cm 0cm 0mm,
width=2\columnwidth]{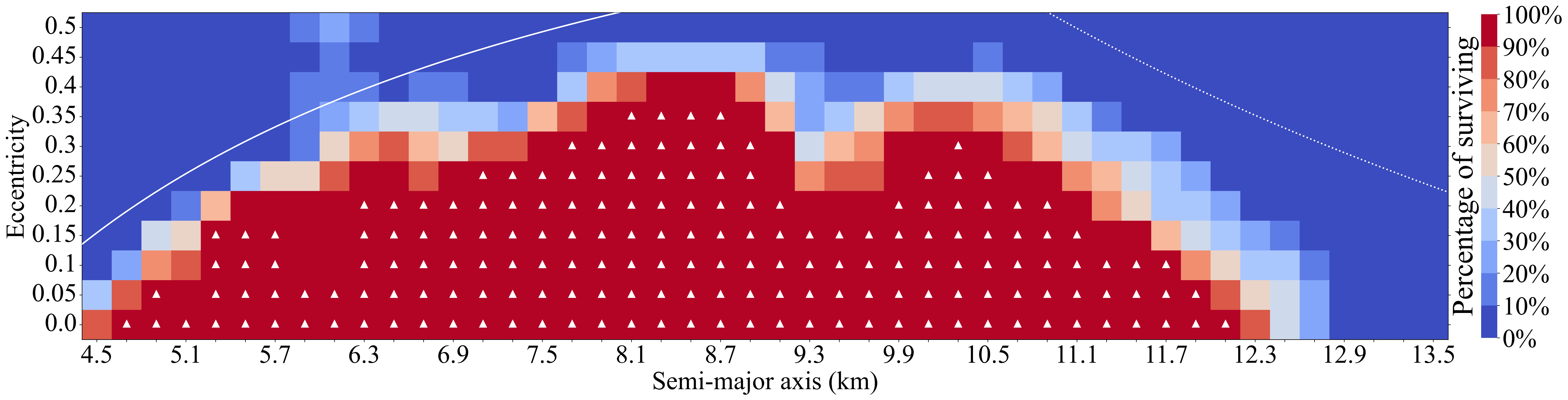}\label{bfig:Reg2_SemSrp_Retro}}
\end{center}
\caption{\label{fig:Reg2_SemSrp_Retro} Diagram of stability for region 2 for two years considering retrograde trajectories. (a) it is a reproduction of the bottom figure 4 from the standard case \citep{Araujo2015} and (b) shows the system considering the irregular shapes of Alpha and Beta. Each box represents a set of 100 particles and the colorbar represents the percentage of surviving particles. The white triangles indicate 100\% of survival. The lines indicate the collision-lines with Gamma (full line), and Beta (dotted line).}
\end{figure*}

For the mascons retrograde case, some regions were noticeably more disturbed than the standard case. 
The larger values of the semi-major axis became null for survivors due to the elongated shape of Beta 
(Fig. \ref{fig:Reg2_SemSrp_Retro}b), decreasing the initial semi-major axis stability border by about 800 m in comparison with the standard case 
(Fig. \ref{fig:Reg2_SemSrp_Retro}a). The ($a, e$) pairs with large eccentricities near Gamma 
and Beta collision lines did not reach a large percentage of survivors due to the instability caused by them, similar to the standard case. 
A prominent valley near the semi-major axis at 9.3 km was formed, and a minor one was also formed near 7.1 km (Fig. \ref{fig:Reg2_SemSrp_Retro}b). 
The major valley is associated with a 2:1 mean motion resonance with Beta and a 1:4 with  Gamma. 
The combination of these resonances generated an instability that  affected nearby particles 
with an initial eccentricity larger or equal to 0.2. In the minor valley exists a 3:1 
 mean motion resonance with Beta. 
 It is interesting to note that the results of the standard case (Fig. \ref{fig:Reg2_SemSrp_Retro}a) were also found two valleys, but they were at different locations. Consequently, they were associated with resonances different from the ones described above.   


\subsection{Region 3}
\label{region3}

\begin{figure}
\begin{center}
\subfloat[]{\includegraphics*[trim = 0mm 0cm 0cm 0mm,
width=1\columnwidth]{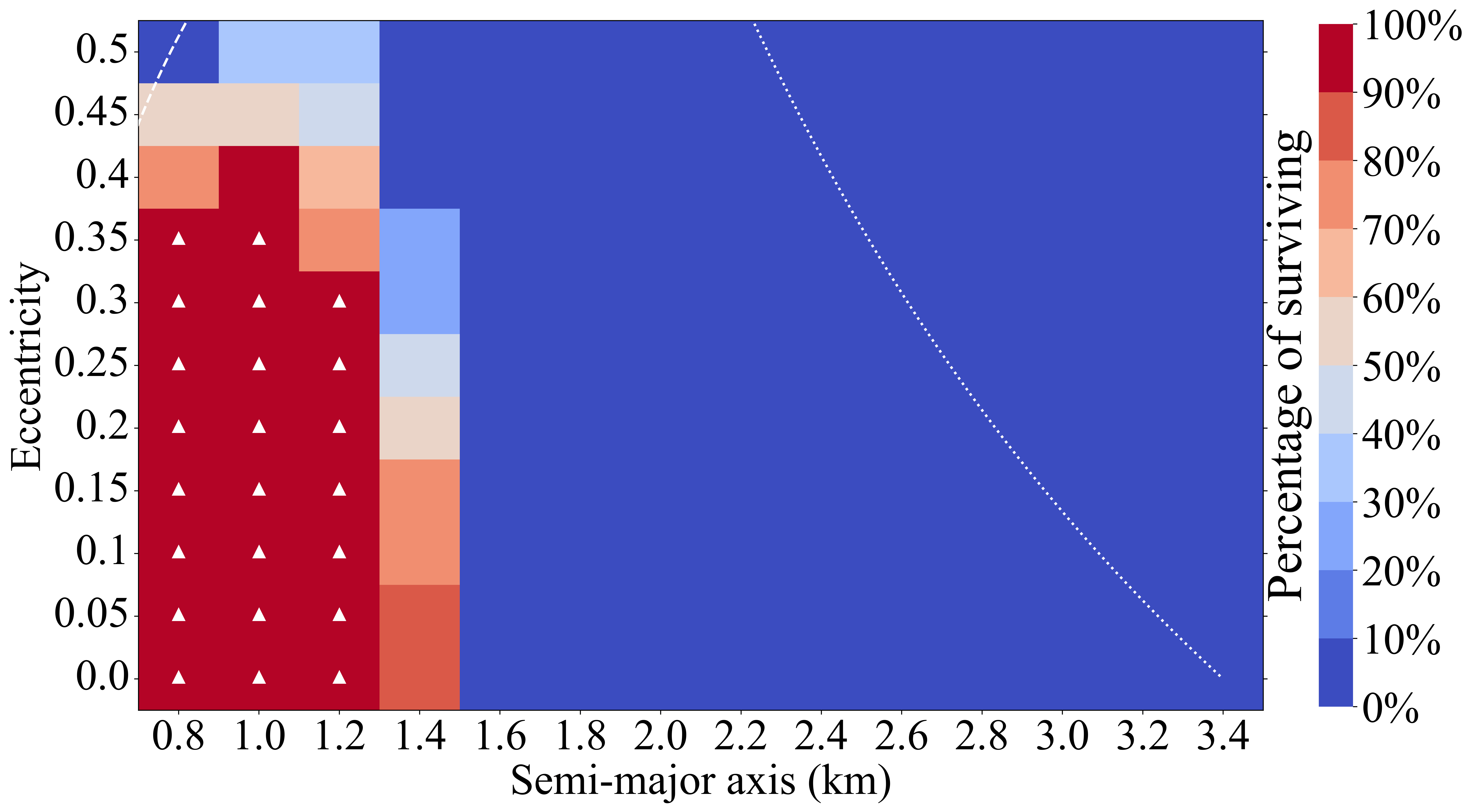}\label{afig:Reg3_SemSrp}}
\end{center}
\caption{\label{fig:Reg3_SemSrp} Reproduction of the diagram of stability for region 3 for two years considering prograde trajectories. The diagram is a reproduction of  figure 4a from the standard case \citep{Araujo2012}. Each box represents a set of 100 particles and the colorbar represents the percentage of surviving particles. The white triangles indicate 100\% of survival. The dashed line indicates the collision-line with Beta for a radius of 543 m. The dotted line corresponds to the Hill's radius of Beta with respect to Alpha and represents the ejection distance.}
\end{figure}

We set 15,400 initial conditions considering the range of $a$ and $e$. The particles are
orbiting Beta, and their motion is restricted by the collision-line of Beta and the 
distance of one Hill’s radii of Beta with respect to Alpha (3.4 km). Figure \ref{fig:Reg3_SemSrp} shows the stability
diagram for the prograde case for our reproduction of figure 4a from \citet{Araujo2012}. 
We obtained only five survived particles for the prograde case system with the Beta's shape model and Alpha as a spherical object. In contrast, for the standard case, most of the particles with a semi-major axis smaller than 1.4 km survive.
Since five survived particles are not a considerable number for stability and considering that this case was already
unstable without  Alpha's shape model and the Sun tide, we did not explore their contribution to the stability of region 3. 
Unlike the standard case, we computed a more significant number of collided particles (about 57\%) and ejected particles
(approximately 43\%). In addition, the particles near the Beta's surface were ejected or collided quickly, with about 78\% of the
collisions and 69\% of the ejections in the first seven days. Thus, the irregularities of Beta's shape cause extreme
instability in the region.

\begin{figure}
\begin{center}
\subfloat[]{\includegraphics*[trim = 0mm 0cm 0cm 0mm,
width=1\columnwidth]{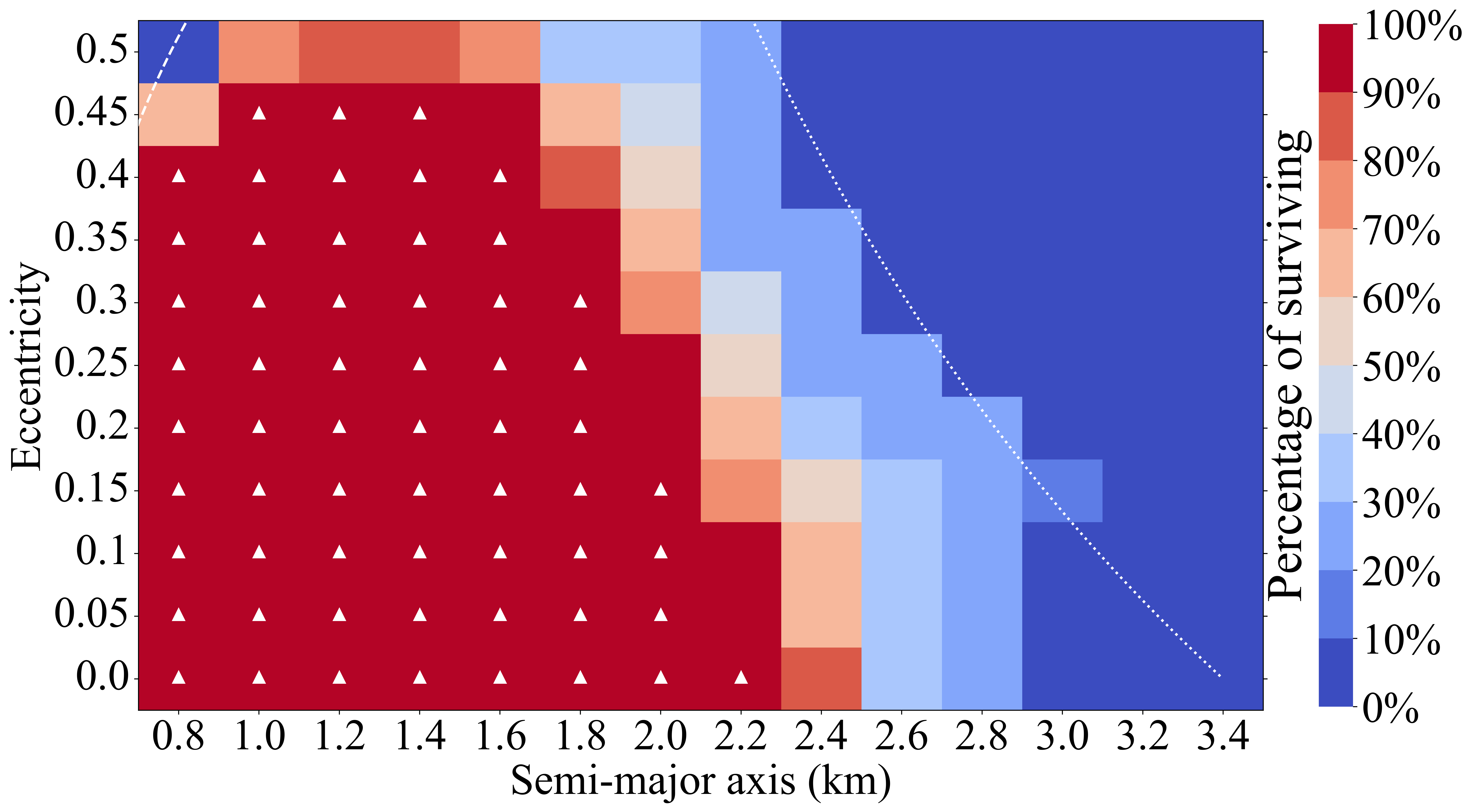}\label{afig:Reg3_SemSrp_Retro}}\\
\subfloat[]{\includegraphics*[trim = 0mm 0cm 0cm 0mm,
width=1\columnwidth]{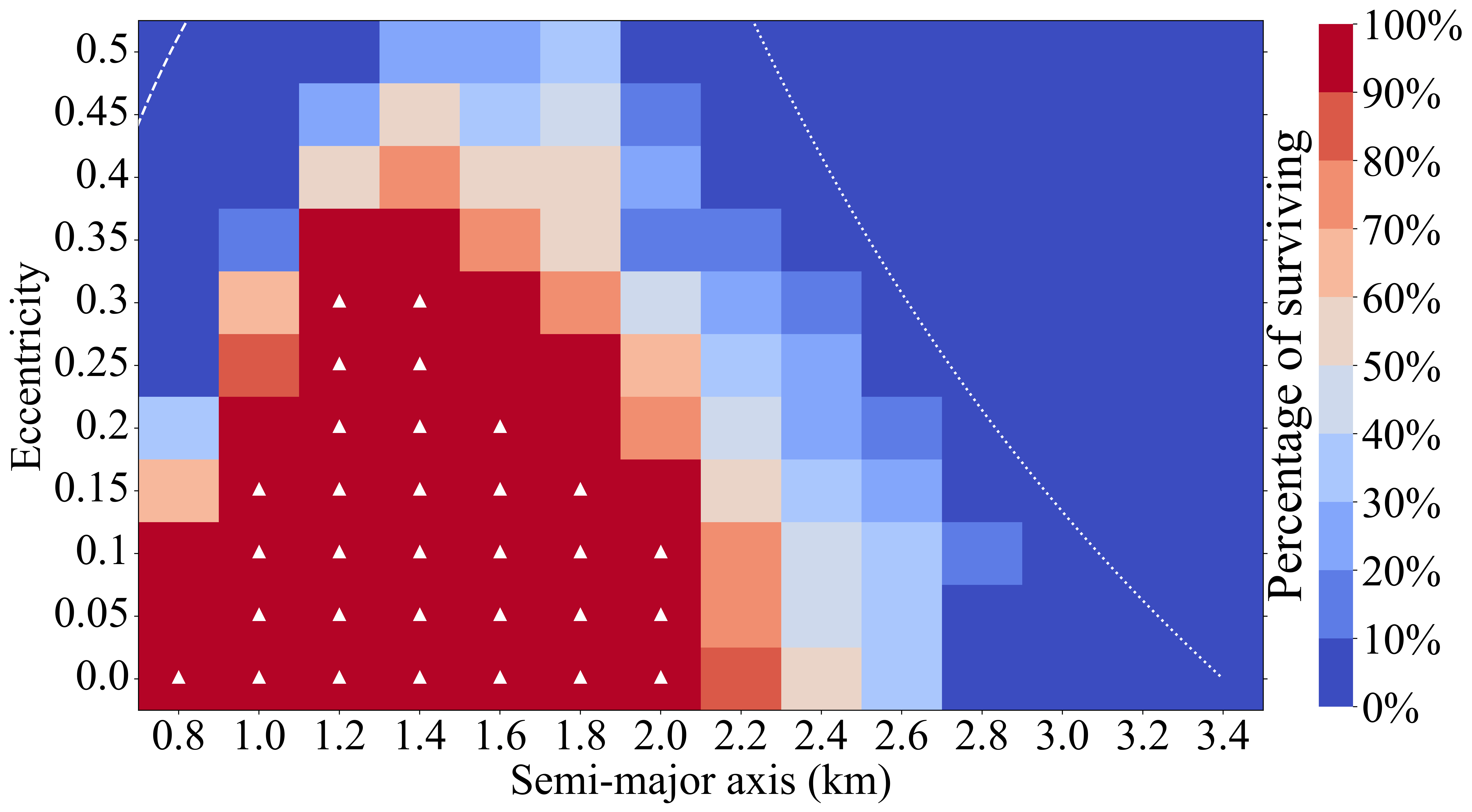}\label{bfig:Reg3_SemSrp_Retro}}\\
\subfloat[]{\includegraphics*[trim = 0mm 0cm 0cm 0mm,
width=1\columnwidth]{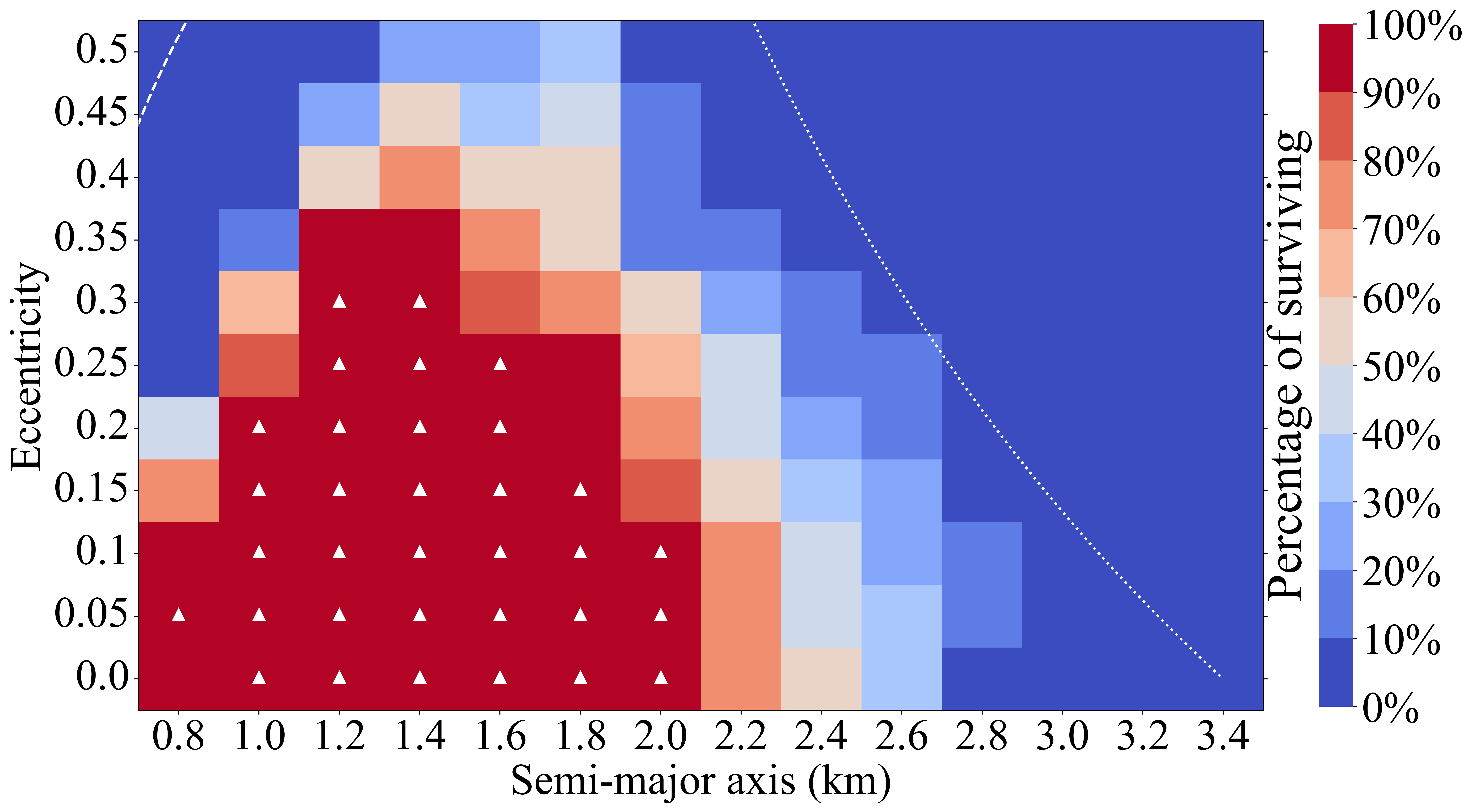}\label{cfig:Reg3_SemSrp_Retro}}
\end{center}
\caption{\label{fig:Reg3_SemSrp_Retro} Diagram of stability for region 3 for two years considering retrograde trajectories. (a) it is a reproduction of figure 6f from the standard case \citep{Araujo2015} and (b) shows the system considering the irregular shape of Beta and (c) considering the irregular shapes of Alpha and Beta.Each box represents a set of 100 particles and the colorbar represents the percentage of surviving particles. The white triangles indicate 100\% of survival. The dashed line indicates the collision-line with Beta for a radius of 543 m. The dotted line corresponds to the Hill's radius of Beta with respect to Alpha and represents the ejection distance.}
\end{figure}

For our mascons case, considering Beta's shape model and Alpha as a spherical body, we detected a decrease of 26.9\% of survived particles (Fig. \ref{fig:Reg3_SemSrp_Retro}b) for the standard case (Fig. \ref{fig:Reg3_SemSrp_Retro}a).
The decrease of survivors occurred at the lower semi-major axis and larger eccentricities.
The apocentre of the particles with a low semi-major axis and large eccentricities reach distances
closer to Beta, giving rise to instability and, eventually, collisions and ejections.
The collisions were about 38\% and 21.7\% for the ejections, where 57\% of these ejections 
and collisions occurred in the first seven days. When we added Alpha's irregular shape model, we did not observe any significant changes (Fig. \ref{fig:Reg3_SemSrp_Retro}c). The statistics of survivors, collisions, and ejections were the same for both scenarios 
(with Alpha as a spherical body and with its irregular shape).

\begin{figure}
\begin{center}
\subfloat{\includegraphics*[trim = 0mm 0cm 0mm 0mm, width=0.9\columnwidth]{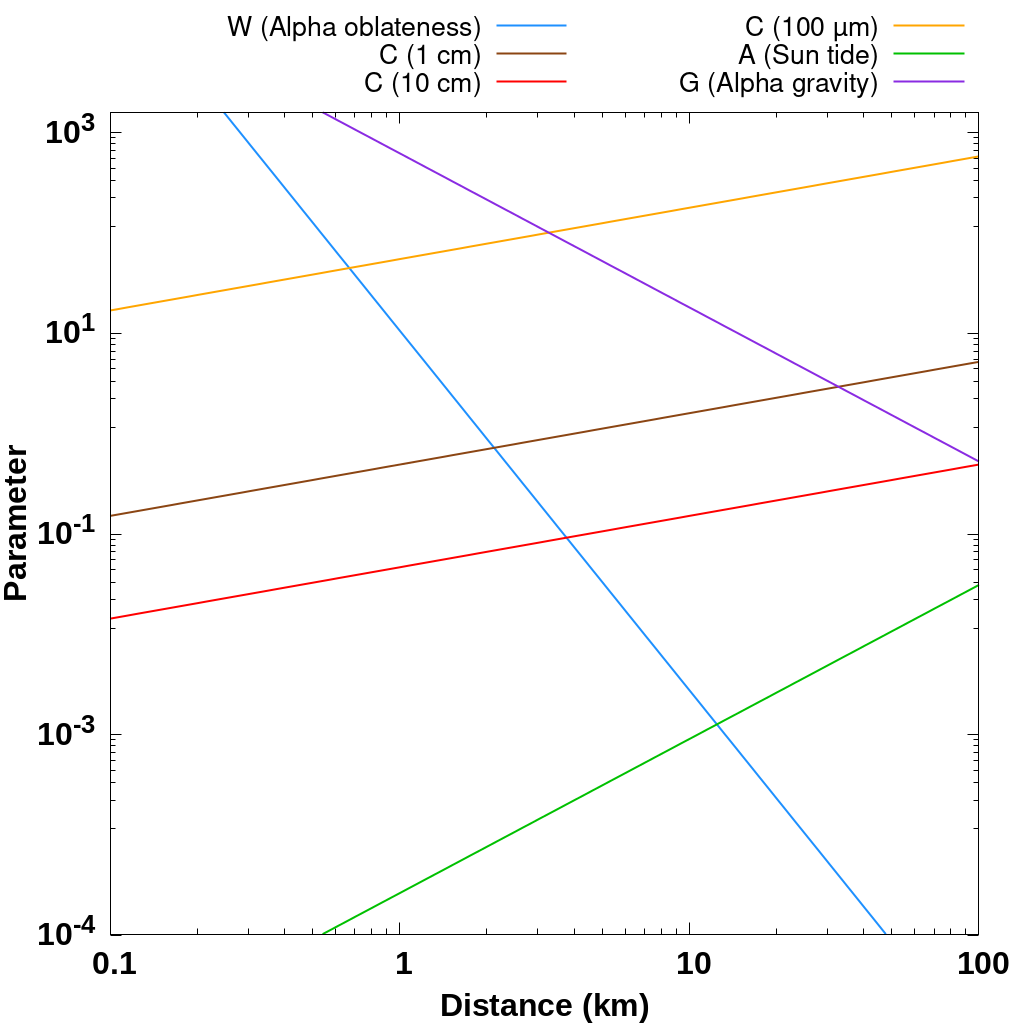}}
\end{center}
\caption{Variation of the dimensionless parameters Sun tide (green), the solar radiation pressure for a particle with a radius of 100 $\mu$m (orange), 1 cm (brown) and 10 cm (red), Alpha's oblateness and gravity (blue and purple, respectively).}
\label{fig:Force}
\end{figure}

Figure \ref{fig:Force} presents the dimensionless parameters 
(see \citet{hamilton1996circumplanetary, moura2020dynamical} for more details) allowing us to
compare the relative strength of each force (solar radiation pressure, Sun`s tide, Alpha's gravity, and oblateness) according to the distance from Alpha. The Sun tide is represented by the color
green and Alpha's gravity and oblateness by the colors purple and blue, respectively. 
The colors orange, brown, and red represent the solar radiation pressure for a particle with a 
density of 1.0 g$\cdot$cm$^{-3}$ and radius of 100 $\mu$m, 1 cm and 10 cm, respectively. 
Since the Alpha's oblateness in region 3 is about the same order of
magnitude as the Sun tide (see Fig. \ref{fig:Force}),
we concluded that the Sun's tidal perturbation may lower affect the region. 
Thus, we did not take into account this perturbation for the third region.

\section{Instability due to the solar radiation pressure}
\label{srp}

Seeking a more realistic approach, we investigated how the solar radiation force changes the stability regions.
The solar
radiation pressure may cause an increase in the particle eccentricity and may change the
stability presented previously. We performed numerical simulations changing the particle 
radius and preserving its density of 1.0 g$\cdot$cm$^{-3}$ (similar to the density of Alpha 
and Beta) to verify the size of the particles that can survive 
nearby the system. Thus, considering an area to mass ratio $A$, $\vec{R}$ as the vector
Sun-particle and $R$ its module, the solar radiation pressure perturbation is given by \citep{scheeres2002spacecraft}

\begin{equation}
\vec{a}_{srp} = \frac{AG^{*}(1+\eta)}{R^3}\vec{R},
 \label{eq: acel_medium}  
\end{equation}
where $G^*=1\times10^{17}$ kg$\cdot$m$\cdot$s$^{-2}$ is the solar constant in function of
the solar luminosity and speed of light, and $\eta$ is the reflectance of the particle. We
consider a totally reflective material, thus $\eta =$ 1. This work do not consider
the shadowing effects, since, compared to the solar radiation pressure and the irregular
gravitational perturbation, they are not significant to create a noticeable change in the
survivors percentage.

Similarly to the section \ref{nearby}, we used the \textsc{N-BoM} package to perform the simulations, but considering the solar radiation pressure \citep{valvano2022apophis}.
We considered the same initial conditions for the three regions.
We also choose a discrete distribution of the radius for the particles to start
in the order of centimeters (e.g., 10, 30, 50, 100, and 150 cm). We opted for a minimum radius 
of 10 cm due to the analyses of the dimensionless parameters of Fig. \ref{fig:Force}. 
As one can see, the perturbation of the radiation pressure on a particle smaller than 10
cm is high enough to generate instability and decrease the percentage of survivors. Thus, we simulated the system for two years considering the retrograde case and our discrete
particles radius. 


\subsection{Region 1}
\label{region1srp}

The irregularity of the central body causes considerable instability in region 1 compared to the
standard case. For the total integration time of the prograde simulation, none 
of the particles has survived, and most of them have collided or ejected in just a few 
days (see section \ref{region1}). Thus, we did not perform a simulation with the solar radiation pressure for the 
prograde case since this perturbation will not generate any stability for the system. 
So, different from \citet{sanchez2019searching} results that found a stable region 
between Alpha and Gamma, when Alpha was modeled as a triaxial ellipsoid and Gamma as
a spherical body, our results showed that region 1 was already unstable without the solar radiation pressure.

\begin{figure}
\begin{center}
\subfloat[]{\vspace{3cm}\includegraphics*[trim = 0cm 0cm 0cm 0mm,
width=.9\columnwidth]{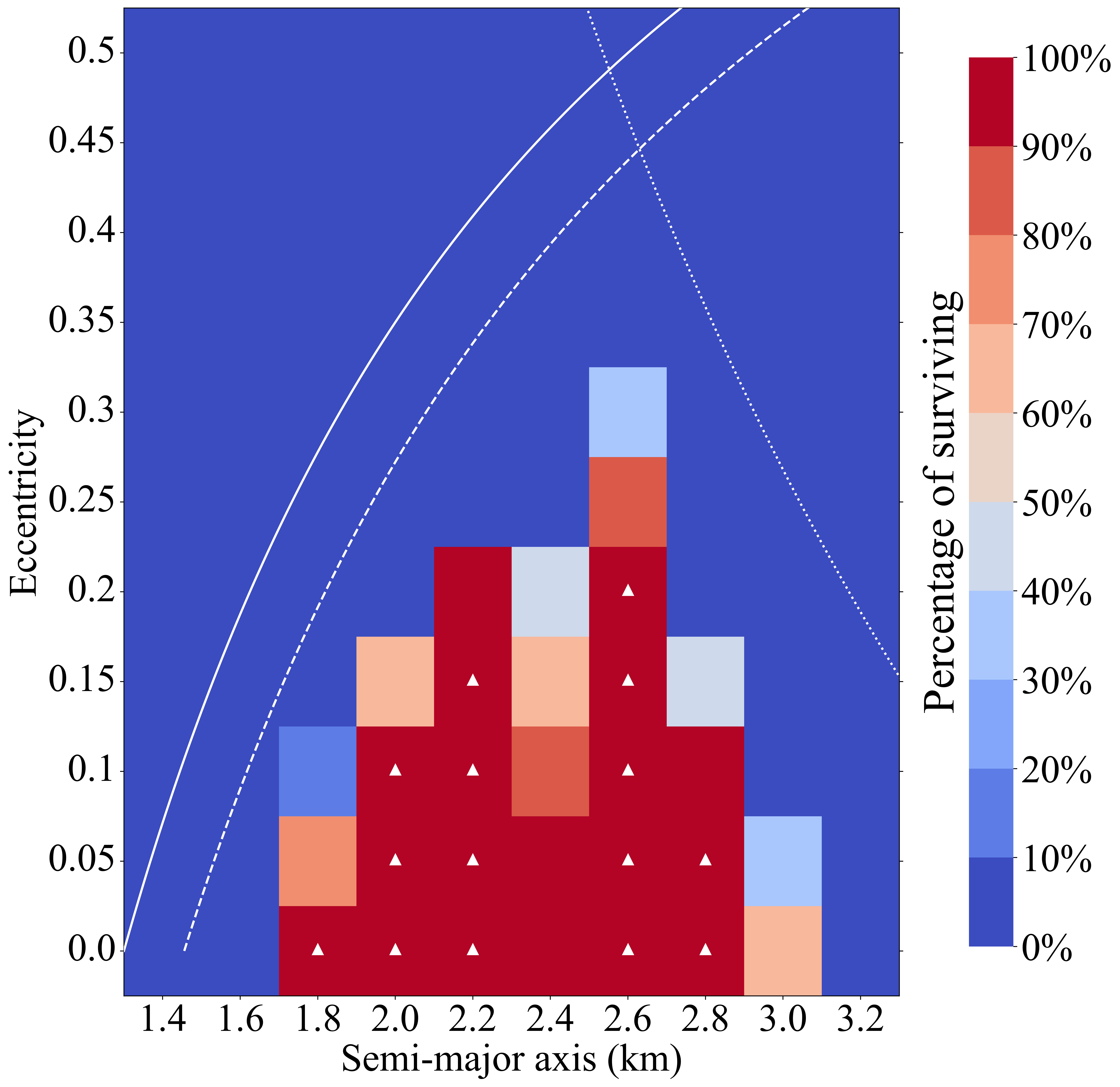}\label{bfig:Reg1_SemSrp_Retro_srp}}
\end{center}
\caption{\label{fig:Reg1_SemSrp_Retro_srp} Diagram of stability for region 1 for two years
considering retrograde trajectories for particles with 10 cm of radius and 
the perturbation of solar radiation pressure. 
The irregular shape of Alpha was considered for this system. Each box represents 
a set of 100 particles and the color bar represents the percentage of surviving particles.
The white triangles indicate 100\% of survival. The lines indicate the collision-lines 
with Alpha (full line for a radius of 1.3 km, dashed line for a radius of 1.46 km), 
and Gamma (dotted line corresponding to the ejection distance).}
\end{figure}

Figure \ref{fig:Reg1_SemSrp_Retro_srp} shows the diagram of stability for region 2 
considering the solar radiation pressure for the retrograde case for a particle with 
10 cm of radius. The percentage of survivors for the simulation was 23\% and 15
pairs of ($a, e$) have 100\% of survival. The distribution of the percentage of survivors
for the boxes is deeply similar to the case without the solar radiation pressure. We could see slight differences in percentage in specific boxes, but nothing significant. 
The gap in the semi-major axis of 2.4 km is still present and the 2:1 mean motion resonance with Gamma.


\subsection{Region 2}
\label{region2srp}

\begin{figure*}
\begin{center}
\subfloat[]{\includegraphics*[trim = 0mm 0cm 0cm 0mm,
width=2\columnwidth]{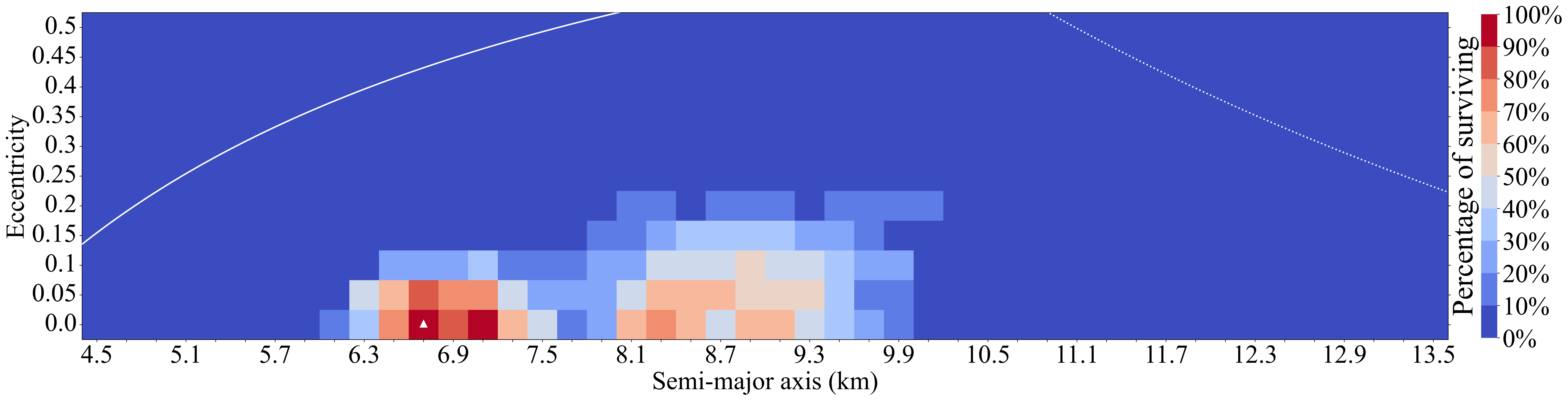}\label{afig:Reg2_comSrp}}\\
\subfloat[]{\includegraphics*[trim = 0mm 0cm 0cm 0mm,
width=2\columnwidth]{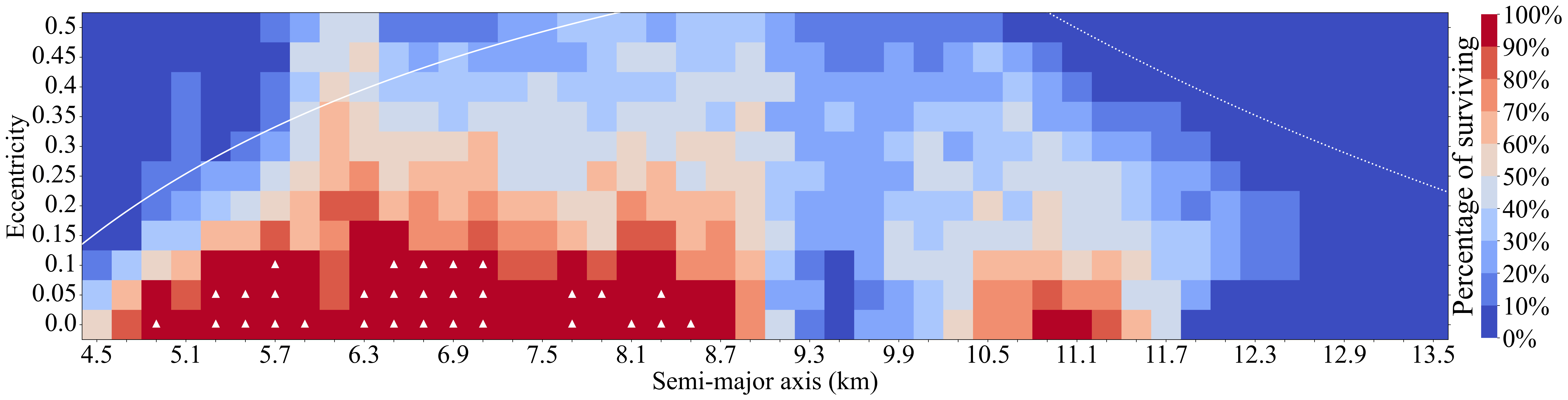}\label{bfig:Reg2_comSrp}}
\end{center}
\caption{\label{fig:Reg2_comSrp} Diagram of stability for region 2 for two years considering (a) prograde and (b) retrograde trajectories for particles with radius of 150 cm and 50 cm, respectively. The irregular shapes of Alpha and Beta were considered for this system. Each box represents a set of 100 particles and the colorbar represents the percentage of surviving particles. The white triangles indicate 100\% of survival. The lines indicate the collision-lines with Gamma (full line), and Beta (dotted line).}
\end{figure*}

If we consider a particle with a radius of 10 cm for the prograde case, no particle 
survived the effects of the solar radiation pressure for two years. Although, when a 
radius of 30 cm was considered, about 2\% of the particles survived. However, no ($a, e$) 
pair presented 100\% stability. Thus, we increased the radius of the particles until 
we reached pairs with 100\% of stability. Increasing the radius to 150 cm,  we obtained a  ($a, e$)
pair with 100\% of stability and similar behavior to the outcomes without the perturbation 
caused by the radiation (Fig. \ref{fig:Reg2_comSrp}a). There are only two main regions 
of surviving particles, and the ($a, e$) pair with 100\% of survival was at the (6.7 km, 0.0) 
pair. Note that the size of the stability region was similar to the case without the 
radiation pressure, and the pair that reached 100\% of survival decreased by 200 m in the initial
semi-major axis value (see Fig. \ref{fig:Reg2_SemSrp}b). The resonances near
the semi-major axis of 7.7 km and 8.1 km are still present.

For the retrograde case, considering particles with the minimum radius (10 cm) 
about 2\% of the particles survived the solar radiation perturbation. Increasing to 30
cm, about 13\% of the particles survived the entire simulation, but no ($a, e$) pair 
presented 100\% of survivors. Thus, we increased the radius of the particles to 50 cm. When 
this radius was considered, about 37\% of the particles survived, and boxes with 100\% 
of survival appeared (Fig. \ref{fig:Reg2_comSrp}b). The location of the previous larger 
valley is now a gap that separates the regions of stability into two. The region of 
stability near Gamma (left side from the gap) was the larger one and presented 30
pairs of ($a, e$) with 100\% of survival. Conversely, the second region of stability
was smaller and presented no boxes with 100\% of survival. The resonances 
in the region of the previous larger valley are still observed. They appeared to be 
more effective in destabilizing the region with the addition of solar radiation 
pressure. We concluded the same for the 3:1   mean motion resonance with 
Beta as well as the resonance near the semi-major axis of 6.1 km. The 1:3  mean motion resonance with Gamma was also preserved but did not create 
a delimited valley as the previously mentioned resonances.


\subsection{Region 3}
\label{region3srp}

\begin{figure}
\begin{center}
\subfloat[]{\includegraphics*[trim = 0mm 0cm 0cm 0mm,
width=1\columnwidth]{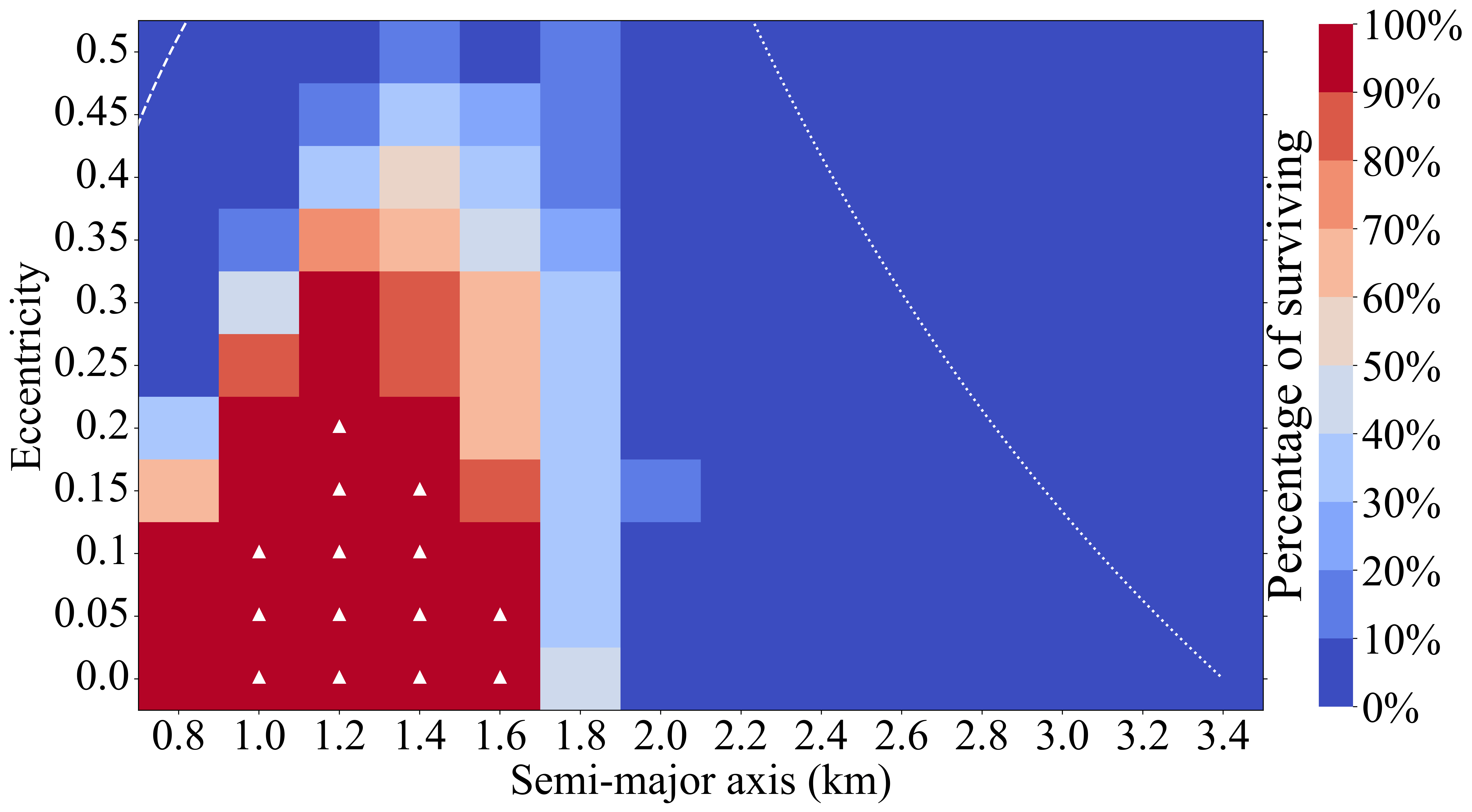}\label{afig:Reg3_comSrp}}
\end{center}
\caption{\label{fig:Reg3_comSrp} Diagram of stability for region 2 for two years considering retrograde trajectories for particles with radius of 30 cm. The irregular shape of Beta was considered for this system. Each box represents a set of 100 particles and the colorbar represents the percentage of surviving particles. The white triangles indicate 100\% of survival. The dashed line indicates the collision-line with Beta for a radius of 543 m. The dotted line corresponds to the Hill's radius of Beta with respect to Alpha and represents the ejection distance.}
\end{figure}

The prograde case around Beta was already 
unstable without the solar radiation pressure. Thus, we did not simulate this region with
this additional perturbation since it cannot produce any stability in the 
region. Considering that the effect of Alpha's shape model was not relevant to 
region 3, we did not consider its shape model to perform the simulations with solar 
radiation pressure for the retrograde case. For the initial radius of 10 cm, about 9\% of 
the particles survived the entire simulation, and only one ($a, e$) pair reached 100\% of
survival. The majority of the particles that survived were near Beta and with low
eccentricities, being the box of  100\% of survival with the initial condition of 
(800 m, 0.0). Increasing the radius to 30 cm, the number of survived particles was
about three times larger, and the region of stability almost reached half of the Hill's 
radius. Similar to the case without the solar radiation pressure, larger 
eccentricities did not produce stability. The stables ($a, e$) 
pairs have eccentricities smaller than 0.2. 

\section{Long-period simulations }
\label{longo}

Considering the stability of a spacecraft, such as the ASTER mission spacecraft, for example, 
two years of stability in the regions would be sufficient for executing a mission. 
However, two years is a short time for natural materials,
such as fragments and small rocks. Thus, we performed simulations for
2,000 years for the survived particles presented in section \ref{nearby} (without the solar
radiation pressure). Since natural material is in retrograde orbits requires
 an improbable process such as gravitational capture have occurred. Therefore, we did not study the retrograde case for a long time.

Once the prograde regions 1 and 3 were already unstable for two years, we only simulated 
region 2 for 2,000 years to verify its long-term stability. Considering the survivors of 
region 2, about 94\% of them collided or were ejected in the first 90 years, and all of them
became unstable before 2,000 years. Thus, we did not include the solar radiation pressure in the
system since it was already unstable.

\section{Final comments}
\label{final}

 The main goal of the current work was to improve the determination of the stability regions in the triple system 2001 SN$_{263}$ found in the works \citet{Araujo2012, Araujo2015}. 
 The advances were made considering more realistic gravitational perturbations due to the bodies' irregular shapes and taking into account the solar radiation pressure.

For the prograde region 1, the irregular shape of Alpha was sufficient to destabilize
the entire region. 
For the retrograde scenario,
24\% of the particles survived the two years of simulation, which is about
half the survivors of the case without the irregularities of Alpha's shape model \citep{Araujo2015}.
Different from the prograde case of region 1, the prograde scenario of region 2 presented 
about 6.7\% survivors, and only one ($a, e$) pair reached 100\% of survival. A valley separated the regions
of stability in two spots, and in this valley, we found
a 3:1  mean motion resonance with Beta.
The retrograde case presented about 8.5 
times more survivors than the prograde case, and the outcomes
with and without the irregular shape models are slightly different. The Irregular 
shape of Beta causes more instability at the end of the region, and the main valley 
becomes more noticeable. Some resonances were also found in this region. 

Similar to the prograde scenario of region 1, the prograde case of region 3 has no 
stability. This result was obtained by only considering the 
irregular shape of Beta. 
In the same
way, we performed the retrograde case. We detected about 40.3\% of surviving and 
31 pairs of ($a, e$) reached 100\% of survival. Adding the Alpha's shape model, 
the statistic was basically the same. 
Thus, we conclude that since the third region
is far from Alpha. Its irregularities did not significantly affect the particles 
around Beta. 
Considering this, we did not add the Sun tide to the system since its
order magnitude in this region is similar to the Alpha's oblateness.

The solar radiation pressure plays an important role in the stability of particles in the system 
and may produce a significant perturbation in the particles’ evolution. For this 
reason, we searched, for each region, the order of magnitude of the radius of the
particles that could survive this perturbation. The prograde case of regions 1 
and 3 was already unstable without the solar radiation pressure, so we did not 
simulate these scenarios. For retrograde region 1, 23\% of the particle with a radius of 10 cm were able to survive. On the other hand, 
24.4\% of the retrograde particles with a radius of 30 cm survived in region 3. For the 
vast region 2, a particle's radius of at least 50 cm for 
the retrograde case and 150 cm for the prograde case was necessary. However, we identified
surviving for a particle with a radius of 30 cm for the prograde case, but it 
was three times smaller and with no box with 100\% of survival. In short, the only
prograde case scenario capable of surviving was region 2. The
particle radius of the retrograde case was one order of magnitude smaller, and 
the three regions had survivors. Hence, the retrograde scenario was the most stable
and may be a safer choice for a spacecraft.

Since two years are too short for fragments and small rocks to verify stability,
we performed a 2,000 years of simulation. We only considered the prograde case. 
 For regions 1 and 3, we did not simulate as
they were already unstable for two years. Besides, for region 2, 
 none of the particles survived for 2,000 years of integration. Hence 
 natural objects, such as fragments and small rocks, are 
unlikely to exist in regions 1, 2, and 3.

Our discussion about the stability around the three regions of the triple system 
(153591) 2001 SN$_{263}$ might contribute to the planning of the ASTER mission. 
Furthermore, it may help understand the dynamics of triple systems and
perhaps the planning of other future missions.


\section*{Acknowledgements}
This study was financed in part by 
 the Brazilian Federal Agency for Support and Evaluation of Graduate Education (CAPES), in the scope of the Program CAPES-PrInt, process number 88887.310463/2018-00, International Cooperation Project number 3266, Fundação de Amparo à Pesquisa do Estado de São Paulo (FAPESP) - Proc. 2016/24561-0 and Proc. 2019/23963-5, Conselho Nacional de Desenvolvimento Científico e Tecnológico (CNPq) - Proc. 305210/2018-1. RS acknowledges support by the DFG German Research Foundation (project 446102036).

\section*{ORCID iDs}
G. Valvano \orcidicon{0000-0002-7905-1788} \href{https://orcid.org/0000-0002-7905-1788}{https://orcid.org/0000-0002-7905-1788}\\
O. C. Winter \orcidicon{0000-0002-4901-3289} \href{https://orcid.org/0000-0002-4901-3289}{https://orcid.org/0000-0002-4901-3289}\\
R. Sfair \orcidicon{0000-0002-4939-013X} \href{https://orcid.org/0000-0002-4939-013X}{https://orcid.org/0000-0002-4939-013X}\\
R. Machado Oliveira \orcidicon{0000-0002-6875-0508} \href{https://orcid.org/0000-0002-6875-0508}{https://orcid.org/0000-0002-6875-0508}\\
G. Borderes-Motta \orcidicon{0000-0002-4680-8414} \href{https://orcid.org/0000-0002-4680-8414}{https://orcid.org/0000-0002-4680-8414}\\

\section*{Data Availability}
The data underlying this article will be shared upon reasonable request to the corresponding authors.



\bibliographystyle{mnras}
\bibliography{example} 








\bsp	
\label{lastpage}
\end{document}